\documentclass{article}

\usepackage{PRIMEarxiv}

\usepackage[utf8]{inputenc} 
\usepackage[T1]{fontenc}    
\usepackage{hyperref}       
\usepackage{url}            
\usepackage{booktabs}       
\usepackage{amsfonts}       
\usepackage{nicefrac}       
\usepackage{microtype}      
\usepackage{lipsum}
\usepackage{fancyhdr}       
\usepackage{graphicx}       
\graphicspath{{media/}}     
\usepackage{graphicx}%
\usepackage{multirow}%
\usepackage{amsmath,amssymb,amsfonts}%
\usepackage{amsthm}%
\usepackage[title]{appendix}%
\usepackage{xcolor}%
\usepackage{textcomp}%
\usepackage{manyfoot}%
\usepackage{booktabs}%
\usepackage{algorithm}%
\usepackage{algorithmicx}%
\usepackage{algpseudocode}%
\usepackage{listings}%
\usepackage{pdflscape}
\usepackage{array}
\usepackage{blindtext}
\usepackage{pdflscape}
\usepackage{array}
\usepackage{booktabs}
\usepackage{rotating}
\usepackage{pifont}
\usepackage{changepage}
\usepackage{colortbl}
\usepackage{tikz}
\usepackage{adjustbox}
\usepackage{array}
\usepackage{pdflscape}
\usepackage{geometry}

\title{Evaluating Human-AI Collaboration: A Review and Methodological Framework}

\author{
  George Fragiadakis, Christos Diou, George Kousiouris, Mara Nikolaidou \\
  Department of Informatics \& Telematics \\
  Harokopio University of Athens \\
  \\
  \texttt{\{gfragi, cdiou, gkousiou, mara\}@hua.gr} \\
}

\begin{document}
\maketitle

\begin{abstract}
    The use of artificial intelligence (AI) in collaborative working environments, termed Human-AI Collaboration (HAIC), has become vital across diverse domains, enhancing decision-making, efficiency, creativity, and innovation. Despite its wide potential, evaluating HAIC effectiveness poses significant challenges due to the complex, dynamic, and reciprocal interactions between humans and AI systems. This paper provides a comprehensive analysis of existing HAIC evaluation methods and introduces a novel, domain-agnostic framework for systematically assessing these systems. To support structured assessment, our framework incorporates a decision-tree-based approach that helps select relevant metrics tailored to distinct HAIC modes (AI-Centric, Human-Centric, and Symbiotic). By integrating both quantitative and qualitative evaluation criteria, the framework aims to address key gaps in HAIC assessment methodologies. We highlight the unique challenges posed by evaluating creative and linguistic AI applications, such as large language models and generative AI in the arts, underscoring the need for tailored evaluation approaches in these emerging areas. This work lays the foundation for future research and empirical validation, offering a structured methodology to enhance the evaluation of HAIC systems.
\end{abstract}


\section{Introduction}
Evaluating Human-AI Collaboration (HAIC) remains a significant challenge due to the complex interactions between humans and AI systems. The effectiveness of HAIC depends not only on system performance but also on the quality of human-AI interaction, trust, and adaptability. However, existing evaluation methods often fail to capture these multidimensional aspects, making it difficult to establish comprehensive assessment criteria.

To address this issue, a structured framework that provides a foundation for systematically assessing HAIC across various domains, is introduced in the paper. Is is inspired by existing HAIC evaluation methodologies, while it aims at integrating and complement them. While the framework has not yet been applied in real-world scenarios, it is designed to be adaptable and to serve as a reference for future empirical validation.

Section~\ref{sec:literature} presents a literature review on HAIC evaluation methods, summarizing different approaches, key metrics, and domain-specific challenges. This review highlights gaps in existing evaluation strategies, particularly their inability to fully capture the dynamic and reciprocal nature of HAIC interactions. Based on the finding of this review and the shortcoming identified we introduce an structured evaluation framework~\cite{bolman2013reframing} that systematically assesses HAIC across multiple domains (Sections~\ref{sec:framework}–~\ref{sec:implications}) . This framework integrates key evaluation factors such as goals, interaction quality, and task allocation, ensuring a standardized and adaptable assessment tool for practitioners and researchers.

\subsection{Background}
The rapid advancement of Artificial Intelligence (AI) has transformed multiple domains, leading to a shift from traditional Human-Machine Interaction (HMI) toward Human-AI Collaboration (HAIC)~\cite{makarius2020rising}. Unlike conventional HMI, which emphasizes usability and task performance~\cite{parasuraman2000model, iso19989241}, HAIC involves a symbiotic partnership where humans and AI work together to achieve shared goals.

This paradigm shift has demonstrated immense potential across diverse sectors, including manufacturing~\cite{ehsan2022symbiotic, emmanouilidis2021}, financial services~\cite{buckley2021regulating, lui2018artificial}, healthcare~\cite{tschandl2020human, lai2021human}, and education~\cite{holstein2022designing}. However, despite its widespread adoption, evaluating HAIC remains an unresolved challenge. Traditional assessment frameworks, which focus on efficiency and accuracy, are insufficient to capture the complexity of human-AI interaction, adaptability, and trust.

This limitation motivates the need for a comprehensive review of HAIC evaluation methodologies (Section~\ref{sec:literature}) and the development of a structured framework that accounts for these multidimensional aspects (Sections~\ref{sec:framework}–~\ref{sec:implications}).

\subsection{Challenges}
While HAIC offers numerous opportunities, evaluating its effectiveness remains a complex and unresolved challenge. Recent work by Woelfle et al. (2024)~\cite{woelfle2024benchmarking} highlights the limitations of traditional evaluation metrics such as task performance, response time, and user satisfaction. Their study emphasizes the need for more comprehensive benchmarks that account for the dynamic and reciprocal nature of HAIC interactions, incorporating metrics like evidence appraisal quality, trust, and adaptability. Such advancements aim to better reflect the multifaceted and collaborative dimensions of human-AI partnerships.

One of the primary challenges lies in assessing the individual and joint contributions of humans and AI systems to collaborative outcomes. Unlike traditional human-machine paradigms, HAIC involves a deeper integration of human intuition and AI capabilities. Evaluating this synergy requires novel metrics that effectively capture the mutual influences between collaborators.

The unique challenges posed by creative and linguistic domains further highlight the limitations of traditional evaluation methods. Applications such as Large Language Models (LLMs) and generative AI in the arts demonstrate the transformative potential of AI to enhance creativity and human expression. However, evaluating these systems requires specialized metrics that move beyond task-based measures. For instance, assessing the interplay of human and AI contributions in artistic creation or the ethical implications of AI-generated content demands methodologies that capture qualitative and contextual nuances. While this study provides a foundational framework for HAIC evaluation, exploring the specific requirements of creative and linguistic domains is an essential avenue for future research.

Additionally, measuring the quality of human-AI interactions remains a critical yet underexplored area. Interaction quality encompasses shared decision-making processes, communication dynamics, and mutual adaptability. Evaluating the ability of both the AI system and human participants to respond to evolving circumstances is crucial for understanding the resilience and effectiveness of HAIC systems in dynamic and high-stakes environments.

\subsection{Proposed Approach}
Drawing inspiration from established frameworks~\cite{rouse1987architecture, parasuraman2000model}, our approach emphasizes ethical considerations, adaptability, and domain-specific customization. The proposed framework provides a comprehensive structure for assessing HAIC across various modes of interaction, from AI-centric to symbiotic collaborations, integrating quantitative and qualitative measures.
Furthermore, it aims to assist researchers and practitioners in identifying existing gaps and opportunities for improving collaboration between humans and AI. By incorporating domain-specific insights, the framework is designed to address practical challenges and applications across fields such as healthcare, finance, and education, thereby demonstrating its versatility and relevance in diverse contexts.

\subsection{Contributions}
This paper makes the following key contributions:
\begin{itemize}
	\item  A review of existing HAIC evaluation methodologies, identifying key gaps and challenges that limit their applicability across diverse domains.
	\item The introducion a structured evaluation framework for HAIC, integrating both quantitative and qualitative metrics to assess performance, interaction quality, and task allocation.
	\item The framework is domain-agnostic, designed to support AI-centric, human-centric, and symbiotic collaboration models, offering a flexible structure applicable to various application areas.
\end{itemize}

\section{Literature Review}\label{sec:literature}
This section first examines the transition from HMI to HAIC, highlighting key evaluation challenges. It then explores the fundamental elements that define HAIC interactions, providing the foundation for developing a structured evaluation framework.

\subsection{From Human-Machine Interaction to Human-AI Collaboration}
The rapid advancement of HAIC necessitates a broader evaluation perspective that goes beyond traditional HMI usability assessments~\cite{nielsen1994usability,shneiderman1983direct,norman2013design}. HMI methods have primarily been designed to measure user experience and task performance, relying on metrics such as response time, accuracy, and efficiency. However, HAIC introduces new challenges that extend beyond interface usability, requiring metrics that capture interaction quality, collaboration effectiveness, and trust.

Okamura and Yamada (2020)~\cite{okamura2020adaptive} emphasize the importance of system transparency in maintaining trust calibration between humans and AI agents. Over-trusting AI can lead to automation bias, while insufficient trust may result in underutilization. Similarly, Yue et al. (2023)~\cite{yue2023impact} examine how different human-AI collaboration models influence user perceptions and adoption, reinforcing the need for evaluation methods that incorporate trust, interpretability, and adaptability.


\subsection{Core Elements and Modes of HAIC}\label{sec:elements}
Human-AI Collaboration (HAIC) is defined by the cooperative partnership between individuals and AI systems to accomplish shared objectives. Understanding its key elements is essential for designing effective evaluation methodologies.

\begin{itemize}
	\item \textbf{Tasks}: HAIC systems support diverse tasks, ranging from automated decision-making to creative problem-solving. The nature of a given task dictates the required level of collaboration between humans and AI~\cite{dellermann2021future,seeber2020machines}. For instance, in healthcare, an AI system may analyze medical images, while a radiologist interprets and integrates results into a final diagnosis.

	\item \textbf{Goals}: Collaboration in HAIC is driven by individual and collective goals. AI objectives often include enhancing efficiency and accuracy, while human goals may involve skill augmentation and contextual reasoning~\cite{crandall2018ishowo}. Continuing with the radiology example, the AI assists in accurate disease detection, while radiologists focus on clinical decision-making and patient-centered care.

	\item \textbf{Interaction}: The effectiveness of HAIC is determined by communication quality and feedback mechanisms~\cite{amershi2019guidelines,schwalbe2024comprehensive}. Misalignment in expectations or miscommunication can lead to ineffective collaboration. For instance, AI-driven customer support chatbots must interpret user queries and refine responses dynamically to ensure effective problem resolution.

	\item \textbf{Task Allocation}: Dynamic task allocation is essential in HAIC, ensuring that responsibilities shift based on real-time conditions and expertise distribution~\cite{dubey2020haco}. For example, in advanced driver assistance systems, AI controls navigation under normal conditions, but the human driver intervenes when unexpected obstacles arise.
\end{itemize}

\begin{figure}[htp!]
	\centering
	\includegraphics[width=0.5\textwidth]{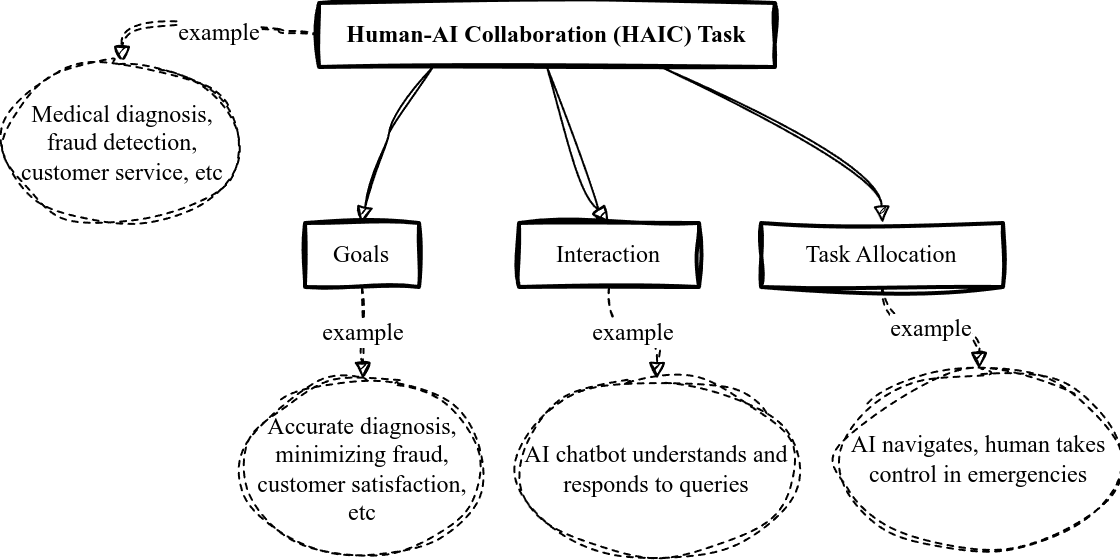}
	\caption{Core elements of Human-AI Collaboration (HAIC) and examples of their application.}
	\label{fig:haic-core-elements}
\end{figure}

Figure~\ref{fig:haic-core-elements} visually represents the key
elements of HAIC, which are important for developing effective evaluation frameworks.
By clearly defining the nature of the collaboration, the intended goals, and the dynamics of interaction,
we can then create metrics and assessments that accurately measure the success and impact of HAIC systems.

One can use any of the HAIC elements, i.e., tasks, goals, types of interaction, and task allocation to differentiate between different HAIC methods. For the purposes of evaluation, however, perhaps the most useful distinction is in terms of allocation~\cite{eigner2024determinants}. We identify three modes, namely AI-centric, Human-centric, and Symbiotic, which are further described below.

\subsubsection*{Human-Centric Mode}
In the Human-centric mode, humans retain the primary decision-making authority, utilizing AI as an augmentative tool to enhance human capabilities without superseding the human role. This mode values human intuition and oversight, with artificial intelligence employed to manage repetitive or data-intensive tasks. It optimizes the unique cognitive abilities of humans for strategic and ethical decision-making while leveraging AI to augment human capabilities.

Supporting this mode, Sutthithatip~\cite{sutthithatip2021explainable} highlights
the use of Explainable AI (XAI) in high-stakes sectors like aerospace to demystify
complex AI operations, thereby enhancing human decision-making. Additionally, How
et al.~\cite{how2020artificial} describe the integration of AI in sustainable
development, which facilitates the involvement of non-experts in data-driven
policy making, illustrating the supportive role of AI in enhancing human capabilities.

\subsubsection*{Symbiotic Mode}
The Symbiotic mode represents a balanced partnership where humans and AI systems collaborate closely, mutually enhancing each other's capabilities. This mode is characterized by a two-way interaction, shared decision-making, and a continuous exchange of feedback, aiming to achieve collective goals through a synergistic relationship. This mode is particularly relevant in complex tasks where both human intuition and AI's computational capabilities are essential for optimal outcomes.

Studies like those by Cesta et al.~\cite{cesta2016towards} explore the symbiotic integration of AI in operational settings, enhancing the collaborative dynamics between humans and robots. Moreover, Jokinen \& Watanabe~\cite{jokinen2019boundary} emphasize the crucial role of trust in these interactions, which is fundamental for a successful symbiotic relationship. Mahmud et al.~\cite{mahmud2023study} and Sowa et al.~\cite{sowa2021cobots} further illustrate how symbiotic HAIC can revolutionize creative fields and managerial roles by combining human creativity with AI's analytical strengths to foster innovation and improve decision-making.

A key example of the symbiotic mode is the \emph{Learning to Defer (L2D)} paradigm,  where AI systems are designed to recognize their own limitations and defer decisions to human judgment when necessary. This approach enhances collaborative decision-making by leveraging the strengths of both humans and AI.
Recent studies have expanded on this concept. Alves et al. (2024)~\cite{alves2024cost} introduced a cost-sensitive L2D framework that considers scenarios with multiple experts and workload constraints, emphasizing the need for flexibility in real-world Human-AI Collaboration (HAIC) systems. Additionally, Hemmer et al. (2022)~\cite{hemmer2022forming} proposed an approach that trains a classification model to complement the capabilities of multiple human experts. 
Their work underscores the adaptability of the HAIC systems' participants, showcasing the ability to dynamically modify task allocation according to the availability and expertise of various agents.

\subsubsection*{AI-centric Mode}
The AI-centric mode designates AI as the primary agent in the collaboration,
where the AI system leads decision-making processes and operates with minimal
human intervention. This mode often features automated interactions where the
AI executes tasks independently, aimed at enhancing system capabilities and
overall efficiency. Such a setup focuses on improving AI functionalities,
maximizing system efficiencies, and generally involves a one-directional
interaction from AI to human, primarily concentrating on tasks suited for
AI's computational abilities.

Recent research underscores the evolving role of AI in traditional human-machine settings. Liu and Zhao~\cite{liu2021role} discuss the allocation of roles in Human-Machine Collaboration (HMC), advocating for optimal strategies that seamlessly integrate intelligent systems into human-centric operations. Similarly, Wang et al.~\cite{wang2020human} examine the shift from human-human to human-AI collaboration, stressing the importance of mutual goal setting and collaborative task management to foster effective partnerships.

In this section, we looked at the fundamentals of Human-AI Collaboration, such as its main components and the various modes of operation. The evaluation of HAIC is critical to ensure that these systems are successful, efficient, and ethical.

The following part looks into the literature on HAIC evaluation, reviewing existing approaches and identifying gaps that suggest areas for further research.

\subsection{Evaluation Approaches in HAIC}\label{sec:approaches}

The evaluation of Human-AI Collaboration (HAIC) is essential to understanding the efficacy of these systems, identifying areas for improvement, and ultimately unlocking their full potential. This section examines current evaluation approaches in HAIC research. While Table~\ref{tab:HAIC_evaluation} provides an  overview of a diverse sample of studies across different domains, the analysis in this section will focus on selected key references to illustrate specific evaluation methodologies and their implications.

Through this analysis of a wide array of studies, our objectives are as follows:
\begin{itemize}
	\item Examine the advantages and drawbacks of current evaluation methodologies.
	\item Identify the distinct difficulties presented by various HAIC applications and domains.
	\item To effectively address the intricate and diverse aspects of HAIC in different sectors.
\end{itemize}

 \begin{table*}[htp!]
			\centering
			\scriptsize
			\caption{Examples of Evaluation of Human-AI Collaboration Across Sectors}
			\rowcolors{2}{gray!10}{white} 
			\begin{tabular}{@{}lp{3cm}p{3cm}ccc>{\raggedright\arraybackslash}p{5cm}@{}}
				\toprule
				\textbf{Reference}                                           & \textbf{Domain}                & \textbf{AI Technology/Method} & \textbf{Qn} & \textbf{Ql} & \textbf{Mix} & \textbf{Additional Info}                                                                                                            \\
				\cmidrule(r){1-7}
				Vollmuth et al. (2023)~\cite{vollmuth2023artificial}         & Healthcare (Oncology)          & Image Analysis                & \ding{70}   &             &              & Accuracy is measured by sensitivity, precision, recall, F1-score. Focuses on human specialists to evaluate AI and HAIC performance.

				\\
				Van Leeuwen et al. (2022)~\cite{van2022does}                 & Healthcare (Radiology)         & Image Analysis                & \ding{70}   &             &              & Emphasizes efficiency using sensitivity, specificity, and outcome metrics.                                                          \\
				Timmons et al. (2023)~\cite{timmons2023call}                 & Healthcare (Mental Health)     & Not specified                 &             & \ding{70}   &              & Investigates bias through interviews and case studies.                                                                              \\
				Farić et al. (2024)~\cite{farivc2024early}                   & Healthcare (Radiology)         & Diagnostic Support Systems    &             & \ding{70}   &              & Assesses early AI use via qualitative feedback from radiologists.                                                                   \\
				Calisto et al. (2022)~\cite{calisto2022breastscreening}      & Healthcare (Breast Cancer)     & Medical Image Analysis        &             & \ding{70}   &              & Focuses on human-AI interaction through qualitative methods.                                                                        \\
				Rezwana et al. (2023)~\cite{rezwana2023designing}            & Design                         & Generative AI                 &             & \ding{70}   &              & Develops interaction design framework using user feedback.                                                                          \\
				Tschandl et al. (2020)~\cite{tschandl2020human}              & Healthcare (Dermatology)       & Image Analysis                & \ding{70}   &             &              & Uses quantitative success metrics in high-stakes environments.                                                                      \\
				Siu et al. (2021)~\cite{siu2021evaluation}                   & Game-Based                     & Reinforcement Learning        & \ding{70}   &             &              & Uses game performance metrics in Hanabi game as a testbed.                                                                          \\
				Reddy et al. (2021)~\cite{reddy2021evaluation}               & Healthcare                     & Not specified                 &             & \ding{70}   &              & Proposes a qualitative framework focusing on ethics and user perspectives.                                                          \\
				Sharma et al. (2023)~\cite{sharma2023human}                  & Mental Health                  & Reinforcement Learning        & \ding{70}   &             &              & Measures conversational empathy improvements through qualitative feedback.                                                          \\
				Yik et al. (2024)~\cite{yik2024neurobench}                   & General (Benchmarking)         & Neuromorphic Computing        & \ding{70}   &             &              & Provides quantitative framework for unique AI context.                                                                              \\
				Rastogi et al. (2023)~\cite{rastogi2023supporting}           & General (Auditing)             & LLM                           &             & \ding{70}   &              & Examines auditing collaboration using qualitative assessments.                                                                      \\
				Zhao et al. (2023)~\cite{zhao_chatspot_2023}                 & Multimodal Interaction         & LLM with Visual Input         & \ding{70}   &             &              & Studies interaction through multimodal input performance metrics.                                                                   \\
				Makarius et al. (2020)~\cite{makarius2020rising}             & General (Business)             & Not specified                 &             & \ding{70}   &              & Emphasizes social factors using qualitative methods.                                                                                \\
				Aleid et al. (2023)~\cite{aleid2023artificial}               & Healthcare (Radiology)         & Image Analysis                & \ding{70}   &             &              & Focuses on accuracy using quantitative measures.                                                                                    \\
				Al-Fatlawi et al. (2023)~\cite{al2023ai}                     & Finance (Fraud)                & Genetic Algorithms            & \ding{70}   &             &              & Evaluates detection rates (true/false positives).                                                                                   \\
				Yang et al. (2022)~\cite{yang2022optimal}                    & General (Teaming)              & Behavior Modeling             & \ding{70}   & \ding{70}   &              & Uses quantitative and qualitative data to model human behavior for better interaction.                                              \\
				Aher et al. (2023)~\cite{aher2023using}                      & Simulation-Based               & Large Language Models         & \ding{70}   &             &              & Simulates human behavior using quantitative metrics.                                                                                \\
				Verheijden \& Funk~\cite{verheijden2023collaborative} (2023) & Design                         & Generative AI                 &             & \ding{70}   &              & Enhances design processes through qualitative co-creation feedback.                                                                 \\
				Hauptman et al. (2023)~\cite{hauptman2023adapt}              & General (Teaming)              & Adaptive AI                   &             & \ding{70}   &              & Investigates adaptability perceptions through qualitative assessments.                                                              \\
				Sankaran et al. (2022)~\cite{sankaran2022modeling}           & Manufacturing                  & Recurrent Neural Networks     & \ding{70}   &             &              & Measures process performance using quantitative metrics.                                                                            \\
				Massaro (2022)~\cite{massaro_multi-level_2022}               & Manufacturing (Safety)         & Artificial Neural Networks    & \ding{70}   &             &              & Evaluates safety and risk through quantitative measures.                                                                            \\
				Oravec (2023)~\cite{oravec2023artificial}                    & Education                      & Large Language Models         &             & \ding{70}   &              & Discusses academic integrity using qualitative methods.                                                                             \\
				El-Zanfaly et al. (2022)~\cite{el-zanfaly_sand_2022}         & Design/Art                     & Physical Interface            &             & \ding{70}   &              & Focuses on creativity and interaction through qualitative feedback.                                                                 \\
				Arias-Rosales (2022)~\cite{arias-rosales_perceived_2022}     & Design/Art                     & Shape Generation              &             &             & \ding{70}    & Combines quantitative shape metrics with qualitative value assessments.                                                             \\
				Dziorny et al. (2022)~\cite{dziorny_clinical_2022}           & Healthcare (ICU)               & Decision Support Systems      &             & \ding{70}   &              & Examines high-stakes challenges using qualitative feedback.                                                                         \\
				Vössing et al. (2022)~\cite{vossing2022designing}            & General (Forecasting)          & LightGBM, Neural Networks     &             &             & \ding{70}    & Investigates AI transparency impact using mixed trust/performance metrics.                                                          \\
				Chakravorti et al. (2022)~\cite{chakravorti2022artificial}   & Finance                        & Prediction Models             &             & \ding{70}   &              & Combines detection rates with qualitative trust assessments for AI markets.                                                         \\
				Huang \& Rust (2022)~\cite{huang2022framework}               & Marketing                      & Machine Learning              &             & \ding{70}   &              & Develops collaborative AI framework using mixed methods.                                                                            \\
				Sachan et al. (2024)~\cite{sachan2024human}                  & Finance                        & Evidential Reasoning          & \ding{70}   &             &              & Reduces decision noise through quantitative measures.                                                                               \\
				Basu et al. (2021)~\cite{basu2021human}                      & Finance (ICO)                  & Linear Regression             &             &             & \ding{70}    & Studies decision-making using mixed quantitative and qualitative data.                                                              \\
				Dikmen \& Burns (2022)~\cite{dikmen2022effects}              & Finance (Peer-to-Peer Lending) & Explainable AI                &             &             & \ding{70}    & Examines trust and performance using mixed domain knowledge assessments.                                                            \\
				Ma et al. (2024)~\cite{ma2024towards}                        & General (Decision-Making)      & Large Language Models         &             &             & \ding{70}    & Proposes deliberative AI using mixed decision-making metrics.                                                                       \\
				Alon-Barkat \& Busuioc (2023)~\cite{alon2023human}           & Public Sector                  & Regression Models             &             &             & \ding{70}    & Explores public sector interactions through mixed methods.                                                                          \\
				Nikolopoulou (2024)~\cite{nikolopoulou2024generative}        & Education                      & Generative AI                 &             &             & \ding{70}    & Assesses generative AI's impact on pedagogy using mixed methods.                                                                    \\
				\bottomrule
			\end{tabular}
			\label{tab:HAIC_evaluation}
	\end{table*}

\subsubsection{Advancements in Tools, Methodologies, and Real-Time Collaboration for HAIC}

Human-AI Collaboration (HAIC) is increasingly evaluated using both traditional and innovative metrics, with an emphasis on ethical concerns, scalability, and adaptability across domains. Recent advancements in tools and methodologies are reshaping the evaluation landscape, offering novel ways to understand and optimize human-AI interactions.

Fischer (2023)~\cite{fischer2023future} envisions the future of human-AI collaboration as a deeply intertwined partnership, where AI systems are designed not only to assist with tasks but also to foster positive emotional and social interactions. His work explores scenarios where AI colleagues are capable of understanding and responding to human emotions, aiming to enhance collaboration by creating a sense of trust and mutual satisfaction. This forward-looking perspective emphasizes the need for evaluation frameworks that address not only technical performance but also the social and emotional dimensions of human-AI interactions. Fischer's work highlights the importance of incorporating human-centered design principles into AI systems to promote effective collaboration and user well-being.

The incorporation of ethical factors into evaluation frameworks is a growing trend, as evidenced by the work of Huang \& Rust (2022)~\cite{huang2022framework} in marketing and Nasir, Khan, \& Bai (2023)~\cite{nasir2024ethical} in healthcare. These frameworks emphasize the importance of transparency (e.g., disclosing the use of AI in decision-making), fairness (e.g., mitigating bias in algorithms), accountability (e.g., establishing clear lines of responsibility for AI-related outcomes), and human-centered design (e.g., prioritizing user needs and values) in AI implementation.  Lase \& Nkosi (2023)~\cite{lase2023human} further advocate for a Human-centric perspective in AI development, stressing the significance of aligning AI systems with user values and ethical principles. They propose using surveys and interviews to assess user perceptions of fairness, trust, and the impact of AI on their work.

Bojic et al. (2023)~\cite{bojic2023hierarchical} highlight the need for standardized and robust evaluation practices, proposing a hierarchical framework to guide the design of reliable human evaluations of AI systems. This framework outlines a step-by-step process for defining evaluation objectives, selecting appropriate metrics (e.g., accuracy, efficiency, user satisfaction), designing experimental protocols, and analyzing data to draw valid conclusions about the effectiveness of HAIC.

Technological advancements further enable real-time monitoring and adaptive collaboration in HAIC. Recent studies, such as Woelfle et al. (2024)~\cite{woelfle2024benchmarking}, emphasize the importance of benchmarking tools that evaluate human-AI collaboration through diverse performance metrics. These metrics extend beyond traditional KPIs like task completion time and error rates, incorporating dimensions such as evidence appraisal quality and the adaptability of AI systems to human inputs. Such tools provide actionable feedback on collaboration efficacy, fostering dynamic adaptability and enhancing the overall effectiveness of HAIC systems.

Interactive machine learning (iML) interfaces, explored by Saha et al. (2023)~\cite{saha2023human}, represent a significant leap in fostering transparency and user engagement. By providing explanations for AI-generated recommendations and allowing users to adjust system parameters, iML interfaces enhance user comprehension, trust, and satisfaction. Saha et al. evaluate these interfaces using a combination of user surveys, interviews, and task-based performance metrics, demonstrating their potential to facilitate effective human-AI partnerships.

Together, these advancements underscore the importance of integrating ethical considerations, real-time analytics, and interactive tools to develop comprehensive evaluation frameworks for HAIC. These tools and methodologies not only enhance our understanding of human-AI collaboration but also pave the way for more adaptive, transparent, and ethical partnerships between humans and AI systems.

\subsubsection{Analysis of Existing Approaches}

In Table~\ref{tab:HAIC_evaluation}, we show a summary of Human-AI Collaboration (HAIC) evaluation practices we examined for this analysis. While quantitative methods remain dominant, there is a growing recognition of the value of qualitative and mixed-methods approaches for capturing the complex interplay between objective performance and subjective experiences.

\subsubsection*{Quantitative Focus: Objectivity and Measurable Outcomes}

Quantitative evaluations, prioritizing objective measures to gauge system performance and efficacy,including the combination of humans and AI, are prevalent in HAIC research. In the medical field, Vollmuth et al. (2023)~\cite{vollmuth2023artificial} and Van Leeuwen et al. (2022)~\cite{van2022does} use tests like sensitivity, specificity, precision, recall, and F1-score to carefully check how well AI-assisted tools in oncology and radiology can make diagnoses. To compare how well AI can find fraud in the financial world, Al-Fatlawi et al. (2023)~\cite{al2023ai} use detection rates (true positives and false positives), and Sachan et al. (2024)~\cite{sachan2024human} use quantitative measures to reduce decision noise in financial underwriting. Manufacturing studies, such as those by Sankaran et al. (2022)~\cite{sankaran2022modeling} and Massaro (2022)~\cite{massaro_multi-level_2022}, also rely heavily on quantitative metrics to assess process performance and safety compliance, respectively. Researchers have developed quantitative frameworks to assess performance in diverse fields such as game-based AI (Siu et al., 2021)~\cite{siu2021evaluation} and neuromorphic computing (Yik et al., 2024)~\cite{yik2024neurobench} based on specific domain-relevant metrics.

While these quantitative approaches provide valuable insights into system performance and measurable outcomes, their focus on objective data often overlooks the nuanced human factors that influence the adoption and success of HAIC systems.

\subsubsection*{Qualitative Insights: Illuminating the Subjective Dimension}

To complement the quantitative focus, a growing number of studies are embracing qualitative methods to delve into the subjective experiences of users and stakeholders. Timmons et al. (2023)~\cite{timmons2023call} investigate biases in mental health AI through interviews and case studies, shedding light on potential ethical concerns and user perceptions. Their study reveals potential ethical concerns, such as the perpetuation of biases and the varying levels of trust users place in AI systems. Rezwana et al. (2023)~\cite{rezwana2023designing} and Verheijden \& Funk (2023)~\cite{rezwana2023designing} use qualitative assessments in the design field to understand the impact of generative AI on creativity and design co-creation processes. Through focus group discussions and in-depth interviews with designers, these studies highlight how generative AI tools influence creative workflows. They discovered that while AI can enhance creative output by providing novel ideas, it also raises concerns about authorship and originality, which are critical to the design profession.

Sharma et al. (2023)~\cite{sharma2023human} look at how conversational empathy improves in mental health applications using qualitative measures. Reddy et al. (2021)~\cite{reddy2021evaluation} suggest a qualitative framework to evaluate the use of AI in healthcare, with a focus on ethical issues and user perspectives. Their framework includes structured interviews with healthcare providers and patients, and thematic analysis to identify key concerns such as data privacy, trust in AI diagnoses, and the perceived impact on the patient-provider relationship. Their study underscores the need for AI systems to align with ethical standards and user expectations to be effectively integrated into healthcare practices.

By capturing the nuanced perspectives of users and stakeholders, these qualitative studies offer a rich understanding of the human factors that shape the adoption and impact of HAIC. However, the subjective nature of qualitative data can pose challenges in terms of generalizability and requires substantial resources for data collection and analysis.

\subsubsection{The Mixed-Methods Promise: Towards a Holistic and Contextualized Evaluation}

The limitations of relying solely on quantitative or qualitative approaches have led to the adoption of mixed-methods evaluations, which integrate objective performance metrics with subjective experiences. This approach provides the flexibility needed to tailor evaluations to specific domains while balancing the rigor of quantitative data with the depth of qualitative insights.
\paragraph{Domain-Specific Applications} In creative domains such as design and art, Arias-Rosales (2022)~\cite{arias-rosales_perceived_2022} combined quantitative shape generation metrics with qualitative assessments of designers’ perceived value. This dual approach highlighted that while AI-generated designs were technically innovative, their subjective aesthetic value varied significantly among designers, underlining the critical role of qualitative judgment in evaluating creative outputs. Similarly, in finance, Chakravorti et al. (2022)~\cite{chakravorti2022artificial} evaluated AI prediction markets by pairing detection rates with trader and analyst perceptions. Their mixed-methods approach revealed that high prediction accuracy did not necessarily translate into trust, exposing concerns about transparency and interpretability.

\paragraph{Generalizable Frameworks} Mixed methods also demonstrate utility across domains. For instance, Yang et al. (2022)~\cite{yang2022optimal} modeled human behavior by combining task completion rates and error frequencies with user interviews, while V\"{o}ssing et al. (2022)~\cite{vossing_designing_2022} integrated system accuracy metrics with trust assessments, uncovering the significant impact of transparency on user acceptance. In marketing, Huang \& Rust (2022)~\cite{huang2022framework} employed a mixed-methods framework that combined performance metrics, such as click-through rates, with ethical and user satisfaction insights. Together, these studies illustrate how mixed methods enable comprehensive evaluations that capture both technical performance and contextual user perceptions.

\subsubsection{Tailored Evaluation for Domain-Specific Challenges}

HAIC's diverse application domains present unique challenges that demand tailored evaluation approaches, reflecting both technical complexities and the distinct priorities of stakeholders. As discussed in the Appendix~\ref{app:llm_genai} of this work, the evaluation of creative and linguistic AI requires tailored methodologies due to its inherent complexities.



\paragraph{Healthcare: Prioritizing Accuracy, Safety, and Ethical Considerations}

In the healthcare, diagnostic accuracy and patient safety are critical. Studies like Vollmuth et al. (2023)~\cite{vollmuth2023artificial} and Van Leeuwen et al. (2022)~\cite{van2022does} use metrics like sensitivity, specificity but  the high-stakes nature of medical decisions require additional qualitative approaches like interviews and focus groups (Farić et al., 2024~\cite{fabri2023disentangling}; Calisto et al., 2022~\cite{calisto2022breastscreening}) to capture user perceptions, identify biases (Timmons et al., 2023~\cite{timmons2023call}), and ensure alignment with ethical principles (Reddy et al., 2021~\cite{reddy2021evaluation}).

\paragraph{Creative Arts: Balancing Creativity, User Experience, and Ethical Impact}

In creative arts, quantitative metrics (e.g. speed, output quality) are complemented byqualitative methods such as interviews and case studies (Rezwana et al., 2023~\cite{rezwana2023designing}; Verheijden \& Funk, 2023~\cite{verheijden2023collaborative}) to asses AI's influence on creativity and collaboration. Ethical concerns regarding bias and the potential undermining  of human creativity, underscore the need for a balanced evaluation~\cite{wu2021ai}.

\paragraph{Finance: Mitigating Risks, Ensuring Trust, and Balancing Quantitative and Qualitative Insights}

In finance quantitative metrics such as detection rates and false positives are crucial for fraud detection (Al-Fatlawi et al., 2023~\cite{al2023ai}). However, understanding AI's impact on decision-making and market stability also requires qualitative insights through surveys and interviews (Chakravorti et al., 2022~\cite{chakravorti2022artificial}). Mixed-methods approaches, provide a comprehensive view of human-AI interaction in financial markets. For a detailed discussion on the challenges specific to creative and linguistic AI, please refer to Appendix~\ref{app:llm_genai}.

In summary, while evaluation approaches vary significantly across  domains, a   comprehensive, mixed-methods framework is still needed to sully assess HAIC's effectiveness and potential.

\subsection{Towards a Unified Evaluation Framework}
While existing HAIC evaluation approaches offer valuable insights, they remain fragmented. A unified framework is needed to integrate objective performance metrics with subjective experiences across diverse domains. Such a framework would support cross-study comparison, enhance generalizability, and standardize evaluation practices.

This framework should adapt to different industries, ensuring AI-assisted tools are both effective and aligned with human needs. In healthcare, it could integrate diagnostic accuracy with user trust; in creative fields, it could assess AI’s impact on artistic processes; and in finance, it could combine risk metrics with transparency evaluations. By addressing methodological gaps, this approach will help harness AI’s potential while ensuring its responsible integration into human workflows.

\section{A Structured Framework for Human-AI Collaboration}\label{sec:framework}

Unlike previous studies that focus on isolated aspects of HAIC evaluation, the proposed framework provides a structured approach by setting measurable goals, clarifying tasks and reporting metrics~\cite{bolman2013reframing}, that integrates multiple evaluation dimensions, organized in three primary factors and specialized subfactors. The primary factors are:

\begin{itemize}
	\item \textbf{Goals} – Evaluates individual and collective objectives within HAIC systems, with key metrics such as learning curves and prediction accuracy~\cite{li_human-ai_2022,vossing2022designing}.
	\item \textbf{Interaction} – Assesses communication mechanisms, feedback processes, adaptability, and trust. Effective communication and transparency are critical for ensuring human-AI trust and usability~\cite{abedin2022designing}.
	\item \textbf{Task Allocation} – Examines role distribution, efficiency, and decision-making processes, ensuring that AI contributions complement human expertise in dynamic environments~\cite{xiong_challenges_2022}.
\end{itemize}

Each factor is further divided into measurable subfactors and associated metrics, allowing for a robust quantitative and qualitative assessment of HAIC effectiveness. To enhance practical applicability, we develop a decision-tree-based evaluation approach. This structured method assists practitioners in selecting relevant evaluation criteria based on the nature of the HAIC system. The decision tree ensures flexibility, allowing for tailored assessments in various domains, including healthcare, finance, manufacturing, and education.

By offering a holistic assessment tool, this framework addresses the key gaps identified in the literature and establishes a standardized methodology for evaluating HAIC in both research and industry applications.

\subsection{Subfactors \& Metrics of the Framework}



Table~\ref{tab:subfactors} summarizes the detailed breakdown of factors to  subfactors and their associated metrics.

  \begin{table*}[htp]
	\label{tab:subfactors}
			\centering
			\caption{Factors and Subfactors of Human-AI Collaboration with Associated Metrics for Evaluation}
			\begin{tabular}{|p{2cm}|p{1.6cm}|p{5cm}|p{8cm}|}
				\hline
				\textbf{Primary Factor} & \textbf{Subfactor}            & \textbf{Human-AI Collaboration}                                                                                     & \textbf{Metrics for Evaluation}                                                                                                                                                                                                                                                                                                         \\
				\hline
				\multirow{2}{*}{\textbf{Goals}}
				                        & Individual Goals              & AI's Goal: Learn Efficiently, Increase Accuracy. Human's Goal: Provide Effective Teaching, Achieve Task Objectives. & * Learning Curve (rate of improvement over time), * Prediction Accuracy (percentage of correct predictions), * Teaching Efficiency (time/resources spent per unit of learning)~\cite{li_human-ai_2022,cakmak2014eliciting, powers2020evaluation}.                                                                                       \\
				\cline{2-4}
				                        & Collective Goals              & Shared Goal: Maximizing the overall performance of the AI system.                                                   & * Overall System Accuracy (percentage of correct outputs), * Objective Fulfillment Rate (percentage of tasks successfully completed)~\cite{vossing2022designing, kerzner2017project}.                                                                                                                                                   \\
				\hline
				\multirow{4}{*}{\textbf{Interaction}}
				                        & Communication Methods         & Query and Response Mechanism.                                                                                       & * Ease of Use (user satisfaction with communication methods), * Clarity of Communication (understandability of queries and responses)~\cite{hinsen2022can, amershi2014power}                                                                                                                                                            \\
				\cline{2-4}
				                        & Feedback Mechanisms           & Human feedback on AI queries, AI adjustments based on feedback.                                                     & * Frequency of Feedback (number of feedback instances), * Feedback Quality (relevance and specificity of feedback), * Feedback Impact (measured change in AI performance due to feedback)~\cite{amershi2014power, dautenhahn2007socially, hinsen2022can}                                                                                \\
				\cline{2-4}
				                        & Adaptability                  & AI's ability to adapt query strategy and learning based on human input.                                             & * Strategy Adaptation Effectiveness (improvement in query selection based on feedback), * Adaptability Score (composite metric of flexibility and responsiveness)~\cite{hinsen2022can, wenskovitch2020interactive, el2022biases}                                                                                                        \\
				\cline{2-4}
				                        & Trust and Safety              & Trust in AI system, ensuring user safety.                                                                           & * Trust Score (level of user trust in AI system)~\cite{papenmeier2022s}, * Safety Incidents (number of safety-related issues)~\cite{magrabi2010analysis}, * Error Reduction Rate (decrease in errors due to AI assistance)~\cite{parasuraman2000model}, * Confidence (level of user trust in AI recommendations)~\cite{zhang2020effect} \\
				\hline
				\multirow{8}{*}{\textbf{Task Allocation}}
				                        & Complementarity               & AI handles data analysis and uncertainty estimation; Humans provide expertise and clarification.                    & * Query Efficiency (amount of information gained per query), * Expertise Utilization (degree to which human expertise is leveraged)~\cite{xiong_challenges_2022}                                                                                                                                                                        \\
				\cline{2-4}
				                        & Flexibility                   & Adjusting query strategy based on interaction dynamics.                                                             & * Adaptability Score (same as in Interaction),  * Dynamic Task Allocation (ability to adjust task distribution based on real-time needs)~\cite{dubey2020haco}                                                                                                                                                                           \\
				\cline{2-4}
				                        & Efficiency                    & Streamlining the learning process through targeted queries.                                                         & * Model Improvement Rate (rate of improvement in AI performance)~\cite{vossing2022designing, xiong_challenges_2022}, * Learning Curve (same as in Goals),  * Resource Utilization (efficiency in using computational resources)~\cite{vossing2022designing}                                                                             \\
				\cline{2-4}
				                        & Responsiveness                & AI's ability to rapidly adjust to new inputs and feedback.                                                          & * Response Time (time taken to respond to queries or feedback), * Real-time Performance (ability to handle time-sensitive tasks)~\cite{hinsen2022can, lopes2023towards}                                                                                                                                                                 \\
				\cline{2-4}
				                        & Collaborative Decision Making & Joint decisions on data labeling and correction of AI's model.                                                      & * Impact of Corrections (measured improvement in system performance through human intervention)~\cite{brady2024developing, zahedi_human-ai_2021}, * Decision Effectiveness (impact of decisions on overall task success)~\cite{schneider2023assessing, mikalef2021artificial}                                                           \\
				\cline{2-4}
				                        & Continuous Learning           & AI's ability to learn from human input and adjust its model.                                                        & * Model Improvement Rate (same as in Efficiency), * Learning Curve (same as in Goals), * Knowledge Retention (ability to retain learned information over time)~\cite{shafiq2022student}                                                                                                                                                 \\
				\cline{2-4}
				                        & Mutual Support                & AI's ability to provide support to humans in the learning process; Humans support by providing information.         & * Error Reduction Rate (decrease in errors due to AI assistance)~\cite{parasuraman2000model}, * Confidence (level of user trust in AI recommendations)~\cite{yang2022user}, * Task Completion Time (with and without AI support)~\cite{murugesan2023study}                                                                              \\
				\cline{2-4}
				                        & Robustness                    & Ensuring AI system's robustness and reliability.                                                                    & * Adversarial Robustness (resilience to adversarial attacks)~\cite{gokhale2022generalized}, * System Reliability (overall system uptime and fault tolerance)~\cite{kuo2000annotated}                                                                                                                                                    \\
				\hline
			\end{tabular}
	\end{table*}

\subsection*{Goals Factor}

The \emph{Goals} factor ensures that the collaboration has a clear direction and that both human and AI efforts are aligned towards shared outcomes. This factor encompasses two key subfactors: \emph{Individual Goals}, which pertain to the specific aims of each participant, and \emph{Collective Goals}, which represent shared objectives that maximize the overall performance of the HAIC system.

Table~\ref{tab:goals_metrics} below summarizes the metrics used to evaluate these subfactors, providing concise definitions, measurement methods, and relevant references to support the evaluation of the \emph{Goals} factor in Human-AI Collaboration (HAIC) systems.

\begin{table*}[htp!]\label{tab:goals_metrics}
	\centering
	\caption{Summary of metrics for the \emph{Goals} factor in Human-AI Collaboration.}
	\begin{tabular}{|p{3cm}|p{3.5cm}|p{8cm}|}
		\hline
		\textbf{Subfactor} & \textbf{Metric}            & \textbf{Definition and Measurement}                                                                                                                                                                                                                                                      \\ \hline
		\multirow{3}{*}{\textbf{Individual Goals}}
		                   & Learning Curve             & Measures the rate of improvement over time, indicating the effectiveness of human teaching and AI learning. This is plotted by graphing AI performance (e.g., accuracy) over iterations or training samples. A steep learning curve suggests effective learning~\cite{li_human-ai_2022}. \\ \cline{2-3}
		                   & Prediction Accuracy        & The proportion of correct predictions made by the AI model out of all predictions. This can be evaluated using metrics like accuracy, F1 score, or area under the ROC curve, depending on the task~\cite{powers2020evaluation}.                                                          \\ \cline{2-3}
		                   & Teaching Efficiency        & The gain in AI performance per unit of teaching effort by the human. This is measured by tracking improvements in AI accuracy or error rate after human input, as explored in studies such as Cakmak and Thomaz~\cite{cakmak2014eliciting}.                                              \\ \hline
		\multirow{2}{*}{\textbf{Collective Goals}}
		                   & Overall System Accuracy    & Calculated as a weighted average of performance metrics, reflecting the relative importance of each task in achieving the overall goal. This metric evaluates the general accuracy of the AI-human collaborative system, tailored to specific objectives~\cite{vossing2022designing}.    \\ \cline{2-3}
		                   & Objective Fulfillment Rate & The percentage of predefined goals successfully achieved within the collaboration. Often used in project management contexts to track progress and adapted to assess HAIC success~\cite{kerzner2017project}.                                                                             \\ \hline
	\end{tabular}

\end{table*}

Schneider et al.~\cite{schneider2023assessing} provide empirical evidence on quality goal rankings, while Wang et al.~\cite{wang2020human} highlight the necessity of goal alignment in collaborative environments. Fiebrink et al.~\cite{fiebrink2011human} illustrate how human input influences AI's learning processes, emphasizing the mutual dependence of individual and collective goals.

\subsection*{Interaction Factor}

The \emph{Interaction} factor underscores the critical communication mechanisms through which humans and AI systems exchange information, provide mutual feedback, and adaptively respond to each other's inputs. Key subfactors within this factor include \emph{Communication Methods}, \emph{Feedback Mechanisms}, \emph{Adaptability}, and \emph{Trust and Safety}. These subfactors capture the essential elements of interaction in Human-AI Collaboration (HAIC), ensuring seamless and effective communication, the incorporation of feedback, and adaptability to dynamic conditions.

Table~\ref{tab:interaction_metrics}  summarizes the metrics used to evaluate the \emph{Interaction} factor, providing concise definitions and measurement methods for each subfactor.

\begin{table*}[htp!]
	\caption{Summary of metrics for the \emph{Interaction} factor in Human-AI Collaboration.}
	\label{tab:interaction_metrics}
	\centering
	\begin{tabular}{|p{4cm}|p{3.5cm}|p{8cm}|}
		\hline
		\textbf{Subfactor} & \textbf{Metric}                   & \textbf{Definition and Measurement}                                                                                                                                                                                                                                                                                                        \\ \hline
		\multirow{2}{=}{\textbf{Communication Methods}}
		                   & Clarity of Communication          & Assesses how understandable the communication is between humans and AI. This can be evaluated using qualitative methods like expert review or user feedback and quantitative measures like the proportion of correctly interpreted queries or time taken to resolve ambiguities~\cite{hoc2000human, el2022biases}.                         \\ \cline{2-3}
		                   & Ease of Use                       & Often measured through Likert scale questionnaires or user interviews to gauge the user's subjective experience with the system.                                                                                                                                                                                                           \\ \hline

		\multirow{3}{=}{\textbf{Feedback Mechanisms}}
		                   & Frequency of Feedback             & Monitors how often users provide feedback, helping developers assess engagement levels and identify areas for improvement~\cite{dehghani2024trustworthy}.                                                                                                                                                                                  \\ \cline{2-3}
		                   & Feedback Quality                  & Measures the relevance and specificity of feedback, either qualitatively through expert reviews or quantitatively by analyzing the correlation between feedback and subsequent AI performance improvements. Hein et al.~\cite{hein2024acceptance} explore feedback acceptance and motivation when provided by AI versus human supervisors. \\ \cline{2-3}
		                   & Feedback Impact                   & Quantifies the tangible effect of feedback on AI behavior or output quality, often measured by improvements in accuracy, precision, or other relevant metrics~\cite{inkpen2023advancing}.                                                                                                                                                  \\ \hline

		\multirow{2}{=}{\textbf{Adaptability}}
		                   & Adaptability Score                & Combines measures of flexibility and responsiveness to assess how quickly and effectively the AI adapts to changes, as studied by El-Assady et al.~\cite{el2022biases}.                                                                                                                                                                    \\ \cline{2-3}
		                   & Strategy Adaptation Effectiveness & Measures improvement in AI's query selection based on human feedback, incorporating insights from semantic interaction research~\cite{wenskovitch2020interactive}.                                                                                                                                                                         \\ \hline

		\multirow{2}{=}{\textbf{Trust and Safety}}
		                   & Trust Score                       & Assesses user trust in the AI system through surveys and trust scales~\cite{lee1994trust}.                                                                                                                                                                                                                                                 \\ \cline{2-3}
		                   & Safety Incidents                  & Tracks the number of safety-related issues encountered to evaluate the reliability and safety of the AI system.                                                                                                                                                                                                                            \\ \hline
	\end{tabular}

\end{table*}

Ethical considerations such as bias, fairness, and transparency are critical to the success of HAIC. Significant research has underscored the complexity of achieving fair and unbiased AI systems. For example, Barocas et al.~\cite{barocas2023fairness} discuss the ethical implications of bias in ML systems, highlighting the need for transparent and accountable AI practices. Holstein et al.~\cite{holstein2019improving} emphasize the importance of fairness in AI design and implementation, advocating for continuous monitoring and evaluation to mitigate bias.

\subsection*{Task Allocation Factor}

Task allocation in HAIC focuses on strategically distributing responsibilities between humans and AI to leverage the strengths of each party and optimize overall performance. Effective task allocation enhances efficiency, minimizes errors, and increases user satisfaction, making it a critical component in the success of HAIC systems~\cite{horvitz1999principles}.

Table~\ref{tab:task_allocation_metrics} summarizes the key subfactors and metrics associated with task allocation, providing concise definitions and measurement methods.

\begin{table*}[htp!]
	\caption{Summary of metrics for the \emph{Task Allocation} factor in Human-AI Collaboration.}
	\label{tab:task_allocation_metrics}
	\centering
	\begin{tabular}{|p{4cm}|p{3.5cm}|p{8cm}|}
		\hline
		\textbf{Subfactor} & \textbf{Metric}         & \textbf{Definition and Measurement}                                                                                                                                                                                                     \\ \hline

		\multirow{2}{=}{\textbf{Complementarity}}
		                   & Query Efficiency        & Indicates the number of queries the AI needs to reach a certain level of accuracy. Fewer queries signify more efficient learning~\cite{xiong_challenges_2022}.                                                                          \\ \cline{2-3}
		                   & Expertise Utilization   & The degree to which human expertise is leveraged. Measured qualitatively through observation or interviews, or quantitatively through metrics like the percentage of decisions where human input is decisive~\cite{seeber2020machines}. \\ \hline

		\multirow{1}{=}{\textbf{Flexibility}}
		                   & Dynamic Task Allocation & The ability to adjust task distribution based on real-time needs. Assessed by how quickly and effectively the system adapts to changes in workload, priorities, or resources~\cite{tsarouchi2017human}.                                 \\ \hline

		\multirow{3}{=}{\textbf{Efficiency}}
		                   & Model Improvement Rate  & The rate of improvement in AI performance. Calculated as the difference in performance between two time points, divided by the time interval. Applied to metrics like accuracy or error rates~\cite{vossing2022designing}.              \\ \cline{2-3}
		                   & Resource Utilization    & The efficiency of computational resource use (e.g., CPU time, memory). Measured as a percentage of utilized resources compared to the total available.                                                                                  \\ \cline{2-3}
		                   & Learning Curve          & The rate of improvement in AI performance over time, as previously defined in the \emph{Goals} section.                                                                                                                                 \\ \hline

		\multirow{3}{=}{\textbf{Responsiveness}}
		                   & Response Time           & The average time taken for AI or human participants to respond to queries or feedback. A lower response time indicates quicker and more effective communication~\cite{hemmer2022factors, hinsen2022can}.                                \\ \cline{2-3}
		                   & Real-time Performance   & The ability to handle time-sensitive tasks, measured by factors like throughput (tasks completed per unit time) and latency (delay in response)~\cite{lopes2023towards}.                                                                \\ \cline{2-3}
		                   & Learning Curve          & The rate of improvement in AI performance over time, as defined earlier.                                                                                                                                                                \\ \hline

		\multirow{2}{=}{\textbf{Collaborative Decision Making}}
		                   & Impact of Corrections   & Measures the improvement in system performance from human intervention. Quantified by changes in accuracy or error rates after corrections~\cite{zahedi_human-ai_2021, brady2024developing}.                                            \\ \cline{2-3}
		                   & Decision Effectiveness  & The impact of decisions on task success. Measured by comparing outcomes of decisions made with and without AI collaboration~\cite{mikalef2021artificial, schneider2023assessing}.                                                       \\ \hline

		\multirow{3}{=}{\textbf{Continuous Learning}}
		                   & Knowledge Retention     & Assesses the AI's ability to retain learned information over time. Evaluated by testing performance after periods of non-use~\cite{yin2021group}.                                                                                       \\ \cline{2-3}
		                   & Model Improvement Rate  & Same as defined in the \emph{Efficiency} section.                                                                                                                                                                                       \\ \cline{2-3}
		                   & Learning Curve          & Same as defined in the \emph{Goals} section.                                                                                                                                                                                            \\ \hline

		\multirow{3}{=}{\textbf{Mutual Support}}
		                   & Error Reduction Rate    & Measures the decrease in errors made by the AI following human intervention. Demonstrates the effectiveness of human corrections~\cite{parasuraman2000model}.                                                                           \\ \cline{2-3}
		                   & Confidence              & The level of user trust in AI recommendations. Measured using Likert-scale questionnaires or surveys~\cite{lee1994trust}.                                                                                                               \\ \cline{2-3}
		                   & Task Completion Time    & Compares task completion times with and without AI support. Helps assess efficiency gains from HAIC~\cite{murugesan2023study}.                                                                                                          \\ \hline

		\multirow{2}{=}{\textbf{Robustness}}
		& Adversarial Robustness  & The resilience of the AI system to adversarial attacks. Measured through testing against adversarial examples.                                                                                                                          \\ \cline{2-3}
		                   & System Reliability      & The overall system uptime and fault tolerance. Evaluated through system logs and reliability tests.                                                                                                                                     \\ \hline
	\end{tabular}

\end{table*}

This structured approach, detailing subfactors and metrics, ensures a comprehensive evaluation of HAIC systems, enabling continuous improvement and effective collaboration between humans and AI.

The Table~\ref{tab:metrics} offers definitions and calculations for each metric described in Table~\ref{tab:subfactors},facilitating a full knowledge of the assessment metrics used in the structured framework for Human-AI Collaboration (HAIC). These metrics are essential for quantitatively evaluating the efficacy, efficiency, and adaptability of HAIC systems across multiple domains.

	\begin{table*}[htp!]
			\centering
			\scriptsize
			\caption{Evaluation Metrics for Human-AI Collaboration}
			\rowcolors{2}{gray!10}{white} 
			\label{tab:metrics}
			\begin{tabular}{@{}p{3cm}p{6cm}p{8cm}@{}}
				\toprule
				\textbf{Metric}            & \textbf{Formula/Description}                                                                                                         & \textbf{Example}                                                                                                                                                    \\
				\midrule
				Learning Curve             & Graphical representation of performance vs. number of training samples.                                                              & In a medical image analysis task, a graph shows the AI's accuracy in diagnosing tumors steadily increasing over 1000 training images.                               \\
				\addlinespace
				Prediction Accuracy        & Accuracy = $\frac{\text{True Positives} + \text{True Negatives}}{\text{Total Predictions}}$                                          & In a spam filter, 98\% of emails are correctly identified as spam or not spam.                                                                                      \\
				\addlinespace
				Teaching Efficiency        & Efficiency = $\frac{\text{Performance Improvement}}{\text{Time Spent}}$                                                              & A financial advisor spends 2 hours refining an AI model's investment recommendations, resulting in a 10\% improvement in predicted portfolio returns.               \\
				\addlinespace
				Overall System Accuracy    & Overall System Accuracy = $\left(\frac{\text{Number of Correct Outcomes}}{\text{Total Number of Outcomes}}\right) \times 100\%$      & In a manufacturing quality control system, 95 out of 100 products are correctly classified as defective or non-defective, resulting in an overall accuracy of 95\%. \\
				\addlinespace
				Objective Fulfillment Rate & Fulfillment Rate = $\frac{\text{Achieved Objectives}}{\text{Total Objectives}}$                                                      & A collaborative design team using AI achieves 8 out of their 10 project milestones on time, resulting in an 80\% objective fulfillment rate.                        \\
				\addlinespace
				Feedback Impact            & Impact = $\text{Performance Post-Feedback} - \text{Performance Pre-Feedback}$                                                        & A language translation AI's BLEU score improves from 0.65 to 0.72 after incorporating user feedback on mistranslations.                                             \\
				\addlinespace
				Adaptability Score         & Adaptability Score = $\text{Performance Post-Adaptation} - \text{Performance Pre-Adaptation}$                                        & A self-driving car adapts its driving style based on weather conditions, improving safety scores by 12\%.                                                           \\
				\addlinespace
				Query Efficiency           & Efficiency = $\frac{\text{Total Queries}}{\text{Queries to Reach Target Accuracy}}$                                                  & A customer service chatbot reduces the average number of queries needed to resolve a customer issue from 5 to 3.                                                    \\
				\addlinespace
				Error Reduction Rate       & Reduction Rate = $\left(\frac{\text{Errors Before} - \text{Errors After}}{\text{Errors Before}}\right) \times 100\%$                 & In a medical diagnosis task, a doctor corrects 8 out of 10 initial misdiagnoses made by the AI system, resulting in an 80\% error reduction rate.                   \\
				\addlinespace
				Confidence                 & Confidence = $\left(\frac{\text{Correct High Confidence Predictions}}{\text{Total High Confidence Predictions}}\right) \times 100\%$ & A financial AI model correctly predicts stock trends 90 out of 100 times when its confidence level is above 90\%.                                                   \\
				\addlinespace
				Response Time              & Average Time = $\frac{\text{Total Response Time}}{\text{Number of Queries}}$                                                         & An AI-powered search engine returns relevant results in an average of 0.5 seconds per query.                                                                        \\
				\addlinespace
				Model Improvement Rate     & Rate = $\frac{\text{Performance at Time T - Performance at Time T-1}}{\text{Time Interval}}$                                         & A fraud detection AI improves its accuracy from 90\% to 95\% over a period of six months, showing a 5\% annual improvement rate.                                    \\
				\addlinespace
				Resource Utilization       & Resource Utilization = $\left(\frac{\text{Resources Used}}{\text{Total Resources Available}}\right) \times 100\%$                    & An AI video processing system utilizes 60\% of the available GPU capacity while running efficiently.                                                                \\
				\addlinespace
				Impact of Corrections      & Improvement = $\text{Performance Post-Correction} - \text{Performance Pre-Correction}$                                               & Human corrections to an AI writing assistant's suggestions result in a 15\% improvement in the text's readability score.                                            \\
				\addlinespace
				Decision Effectiveness     & Effectiveness = $\left(\frac{\text{Successful Decisions}}{\text{Total Decisions}}\right) \times 100\%$                               & A collaborative medical decision support system, combining AI analysis and physician expertise, results in 95\% of treatment decisions being successful.            \\
				\addlinespace
				Knowledge Retention        & Retention Rate = $\left(\frac{\text{Performance Post-Retention}}{\text{Performance Pre-Retention}}\right) \times 100\%$              & A customer service chatbot maintains 80\% of its accuracy in understanding queries after a period of six months without any further training.                       \\
				\addlinespace
				Task Completion Time       & Time Saved = $\text{Time Without AI} - \text{Time With AI}$                                                                          & An AI-powered manufacturing robot reduces the time to complete a specific task from 10 minutes to 5 minutes.                                                        \\
				\addlinespace
				Trust Score                & Trust Score = $\left(\frac{\text{Trust Ratings from Users}}{\text{Total Trust Scale Maximum}}\right) \times 100\%$                   & A survey shows that users rate their trust in the AI system at 85\% on a 100-point scale.                                                                           \\
				\addlinespace
				Safety Incidents           & Safety Incidents = Number of safety-related issues encountered                                                                       & A self-driving car system logs 2 safety incidents per 1000 miles driven.                                                                                            \\
				\addlinespace
				Adversarial Robustness     & Robustness = $\frac{\text{Performance under Adversarial Conditions}}{\text{Performance under Normal Conditions}}$                    & An AI image recognition system maintains 90\% of its accuracy when tested with adversarial examples.                                                                \\
				\addlinespace
				Domain Generalization      & Generalization = $\frac{\text{Performance across Different Domains}}{\text{Baseline Performance}}$                                   & An AI system trained on financial data achieves 85\% of its accuracy when applied to a healthcare dataset.                                                          \\
				\addlinespace
				System Reliability         & Reliability = $\left(\frac{\text{Uptime}}{\text{Total Time}}\right) \times 100\%$                                                    & An AI server system maintains 99.9\% uptime over a year, indicating high reliability.                                                                               \\
				\bottomrule
			\end{tabular}
	\end{table*}

\subsection{Building the Framework \& Quantifying the Evaluation of HAIC}

\begin{figure*}[htbp]
	\centering
	\includegraphics[scale=0.17]{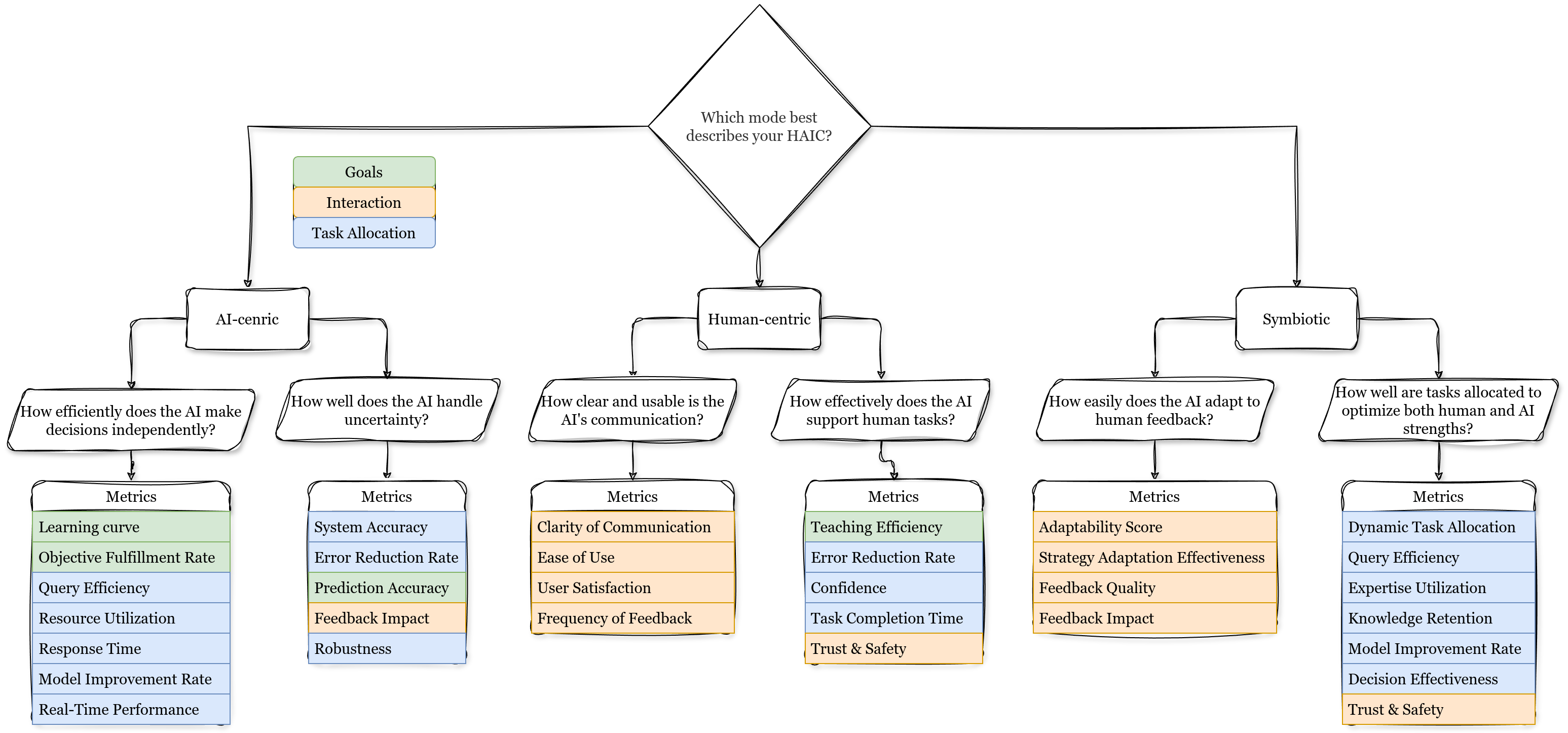}
	\caption{Decision Tree Framework for Evaluating Human-AI Collaboration}
	\label{fig:haic-framework}
\end{figure*}

Building upon influential research by Fischer~\cite{fischer1995rethinking},
Wenskovitch and North~\cite{wenskovitch2020interactive}, and Sharma et.
al~\cite{sharma2023human}, this framework offers a structured approach to
Human-AI Collaboration (HAIC) evaluation. This is achieved using a decision
tree (Fig.~\ref{fig:haic-framework}), that
guides users through a series of questions to identify the most relevant metrics
for their specific HAIC system. This combines the factors, subfactors, and
metrics identified in the previous sections, based on the HAIC modes described
earlier in Section~\ref{sec:elements}, i.e.,
AI-Centric, Human-Centric, and Symbiotic.

Notice in Fig.~\ref{fig:haic-framework}
that each metric has a distinct color based on the main factor that is
coming from the Table~\ref{tab:subfactors}.
This color-coding system helps users quickly identify the primary factor each metric
belongs to, facilitating a more intuitive and user-friendly evaluation process.
The framework is designed to be adaptable and scalable, accommodating a wide range
of HAIC systems and applications. By providing a structured and comprehensive
evaluation approach, this framework aims to enhance the understanding and assessment
of HAIC, fostering the development of more effective and collaborative human-AI systems.

Each metric is color-coded based on its primary factor (Table~\ref{tab:subfactors}), enhancing clarity and usability. This adaptable, scalable framework fosters a comprehensive understanding of HAIC systems, improving their assessment and development.

\paragraph{AI-Centric Collaboration} Focuses on AI autonomy, emphasizing efficiency and uncertainty handling metrics.
\paragraph{Human-Centric Collaboration} Evaluates AI’s communication and support for human tasks, emphasizing usability and user satisfaction.
\paragraph{Symbiotic Collaboration} Assesses adaptability and task distribution to optimize human-AI strengths.

By structuring evaluation through this decision-tree approach, we enhance consistency, enabling better-informed HAIC system development and optimization. This approach simplifies evaluation, tailors assessments to HAIC systems, and remains adaptable as technologies evolve.

\subsection*{Weighting Mechanism \& Overall Score Calculation}
While the decision tree and associated metrics lead the evaluation process, we recognize the importance of a weighting mechanism to provide a more comprehensive evaluation of HAIC effectiveness. A weighting mechanism ensures a comprehensive evaluation, adjusting metric importance based on HAIC mode. Weighted averages determine overall scores, enhancing accuracy and adaptability. Prior research supports this approach, emphasizing context-specific evaluation~\cite{kamar2012combining},~\cite{amershi2014power},~\cite{vossing2022designing}. Future work should refine these weights through empirical studies to strengthen HAIC assessment robustness.

\section{Implications of the HAIC Evaluation Framework to Real-World Domains}\label{sec:implications}

Although it has not yet been empirically validated, the proposed HAIC framework offers theoritical basis for assesing human-AI collaboration across sectors such as manufacturing, healthcare, finance, and education. By emphasizing factors like Adaptability, Confindence and \emph{Clarity of Communication} this framework lays the groundwork for future empirical studies. 

\subsection{Manufacturing}

Manufacturing’s focus on safety, accuracy, and productivity aligns with the HAIC \emph{Symbiotic mode}, which integrates human expertise with AI skills. Studies~\cite{rabby2020modeling, okamura2020adaptive, mohammadi2020mixed, hou2021enabling} demonstrate a shift towards worker-centric, adaptive approaches that highlight critical metrics such as \emph{Adaptability Score, Error Reduction Rate, Confidence}, and \emph{Task Completion Time}.

\subsection{Healthcare}
In medical imaging diagnosis, AI supports human decision-making by enhancing diagnostic accuracy and efficiency~\cite{liu2019comparison, zheng2023overview, kaboudan2023ai, aleid2023artificial, dziorny2022clinical}. The framework targets improvements in diagnostic \emph{Accuracy} and \emph{Efficiency}, emphasizing clear AI outputs. Evaluation parameters include \emph{System Accuracy, Prediction Accuracy, Response Time}, and \emph{Decision Effectiveness}, with further research needed on workflow impacts and patient communication.

\subsection{Finance}
The finance sector benefits from a \emph{Symbiotic} approach where AI’s analytical capabilities complement human expertise to enhance fraud detection and operational efficiency~\cite{atadoga2024ai, al2023ai, jain2023role, ali2022financial}. Key aspects include achieving goals of Accuracy and Efficiency and ensuring clear communication of AI insights. Metrics such as the \emph{Error Reduction Rate} help assess system effectiveness, with future work needed to address ethical and long-term impacts.


\subsection{Education}
In education, a mix of \emph{Symbiotic} and \emph{Human-Centric} models enhances teaching and learning~\cite{ifenthaler2023reciprocal, ji2023systematic, wang2020human, sharma2023human, holstein2022designing, fadiya2023development}. The focus is on improving \emph{Teaching Efficiency} and achieving personalized learning through effective communication. Important metrics include \emph{Task Completion Time} and \emph{Learning Curve}, while further research should explore fostering active student engagement and critical thinking.


\section{Limitations in Investigating Behavioral Factors}\label{sec:limitations}

Our framework for evaluating Human-AI Collaboration (HAIC) emphasizes goals, interaction dynamics, and task allocation while excluding behavioral aspects due to methodological and scope constraints. Investigating behavioral factors involves interdisciplinary approach from psychology, sociology, cognitive science, which require complex and varried methodologies. While these factors are crucial for understanding trust and collaborative goals over time—exemplified by studies such as Yang, Carroll, \& Dragan (2022)~\cite{yang2022optimal}—our study focuses on structural elements that are more readily quantifiable. We advocate for future research to explore behavioral dimensions in-depth to complement our objective framework.

\section{Conclusion}\label{sec:conclusion}
This paper makes a two-fold contribution to HAIC evaluation. First, it offers a comprehensive literature review of the existing methodologies and their limitations. Second, it introduces a structured framework that integrates key evaluation factors identified in the literature and provides a standardized method for assessing HAIC across various domains.

A key gap in current methodologies is the absence of a framework that integrates both qualitative and quantitative metrics across diverse sectors. The framework proposed herein addresses this limitation by offering a structured evaluation approach based on the mode of collaboration rather than a specific sector or AI application. By breaking down evaluation into smaller, manageable components, it provides a flexible method to assess HAIC across different contexts.

Although the framework is theoretical at this stage, future research will focus on empirical validation through software implementation and real-world applications across various sectors, ensuring its adaptability and robustness.
Additionally, our study acknowledges limitations, particularly the exclusion of behavioral and ethical considerations from the evaluation framework. These dimensions, though outside the current scope, are essential aspects of HAIC that should be explored in future research.

\section*{Acknowledgments}
The work leading to these results has received funding from the EU’s Horizon Europe research and innovation programme under GA No. 101120218, project HumAIne. Views and opinions expressed are however those of the author(s) only and do not necessarily reflect those of the European Union. Neither the European Union nor the granting authority can be held responsible for them.

\bibliographystyle{IEEEtran}

\bibliography{references}

\begin{thebibliography}{100}
\providecommand{\url}[1]{#1}
\csname url@samestyle\endcsname
\providecommand{\newblock}{\relax}
\providecommand{\bibinfo}[2]{#2}
\providecommand{\BIBentrySTDinterwordspacing}{\spaceskip=0pt\relax}
\providecommand{\BIBentryALTinterwordstretchfactor}{4}
\providecommand{\BIBentryALTinterwordspacing}{\spaceskip=\fontdimen2\font plus
\BIBentryALTinterwordstretchfactor\fontdimen3\font minus
  \fontdimen4\font\relax}
\providecommand{\BIBforeignlanguage}[2]{{%
\expandafter\ifx\csname l@#1\endcsname\relax
\typeout{** WARNING: IEEEtran.bst: No hyphenation pattern has been}%
\typeout{** loaded for the language `#1'. Using the pattern for}%
\typeout{** the default language instead.}%
\else
\language=\csname l@#1\endcsname
\fi
#2}}
\providecommand{\BIBdecl}{\relax}
\BIBdecl

\bibitem{makarius2020rising}
E.~E. Makarius, D.~Mukherjee, J.~D. Fox, and A.~K. Fox, ``Rising with the
  machines: A sociotechnical framework for bringing artificial intelligence
  into the organization,'' \emph{Journal of Business Research}, vol. 120, pp.
  262--273, 2020.

\bibitem{parasuraman2000model}
R.~Parasuraman, T.~B. Sheridan, and C.~D. Wickens, ``A model for types and
  levels of human interaction with automation,'' \emph{IEEE Transactions on
  systems, man, and cybernetics-Part A: Systems and Humans}, vol.~30, no.~3,
  pp. 286--297, 2000.

\bibitem{iso19989241}
W.~Iso, ``9241-11. ergonomic requirements for office work with visual display
  terminals (vdts),'' \emph{The international organization for
  standardization}, vol.~45, no.~9, 1998.

\bibitem{ehsan2022symbiotic}
E.~T. Esfahani, B.~He, C.-H. Chu, Y.~Liu, R.~Rai, and G.~Ameta, ``Special
  issue: Symbiotic human-ai partnership for next generation factories,''
  \emph{Journal of Computing and Information Science in Engineering}, vol.~22,
  no.~5, 2022.

\bibitem{emmanouilidis2021}
C.~Emmanouilidis, S.~Waschull, J.~A. Bokhorst, and J.~C. Wortmann, ``Human in
  the ai loop in production environments,'' in \emph{Advances in Production
  Management Systems. Artificial Intelligence for Sustainable and Resilient
  Production Systems: IFIP WG 5.7 International Conference, APMS 2021, Nantes,
  France, September 5--9, 2021, Proceedings, Part IV}.\hskip 1em plus 0.5em
  minus 0.4em\relax Springer, 2021, pp. 331--342.

\bibitem{buckley2021regulating}
R.~P. Buckley, D.~A. Zetzsche, D.~W. Arner, and B.~W. Tang, ``Regulating
  artificial intelligence in finance: Putting the human in the loop,''
  \emph{Sydney Law Review, The}, vol.~43, no.~1, pp. 43--81, 2021.

\bibitem{lui2018artificial}
A.~Lui and G.~W. Lamb, ``Artificial intelligence and augmented intelligence
  collaboration: regaining trust and confidence in the financial sector,''
  \emph{Information \& Communications Technology Law}, vol.~27, no.~3, pp.
  267--283, 2018.

\bibitem{tschandl2020human}
P.~Tschandl, C.~Rinner, Z.~Apalla, G.~Argenziano, N.~Codella, A.~Halpern,
  M.~Janda, A.~Lallas, C.~Longo, J.~Malvehy \emph{et~al.}, ``Human--computer
  collaboration for skin cancer recognition,'' \emph{Nature Medicine}, vol.~26,
  no.~8, pp. 1229--1234, 2020.

\bibitem{lai2021human}
Y.~Lai, A.~Kankanhalli, and D.~Ong, ``Human-ai collaboration in healthcare: A
  review and research agenda,'' 2021.

\bibitem{holstein2022designing}
K.~Holstein and V.~Aleven, ``Designing for human--ai complementarity in k-12
  education,'' \emph{AI Magazine}, vol.~43, no.~2, pp. 239--248, 2022.

\bibitem{woelfle2024benchmarking}
T.~Woelfle, J.~Hirt, P.~Janiaud, L.~Kappos, J.~P. Ioannidis, and L.~G. Hemkens,
  ``Benchmarking human--ai collaboration for common evidence appraisal tools,''
  \emph{Journal of Clinical Epidemiology}, vol. 175, p. 111533, 2024.

\bibitem{rouse1987architecture}
W.~B. Rouse, N.~D. Geddes, and R.~E. Curry, ``An architecture for intelligent
  interfaces: Outline of an approach to supporting operators of complex
  systems,'' \emph{Human-computer interaction}, vol.~3, no.~2, pp. 87--122,
  1987.

\bibitem{nielsen1994usability}
J.~Nielsen, \emph{Usability engineering}.\hskip 1em plus 0.5em minus
  0.4em\relax Morgan Kaufmann, 1994.

\bibitem{shneiderman1983direct}
B.~Shneiderman, ``Direct manipulation: A step beyond programming languages,''
  \emph{Computer}, vol.~16, no.~08, pp. 57--69, 1983.

\bibitem{norman2013design}
D.~Norman, \emph{The design of everyday things: Revised and expanded
  edition}.\hskip 1em plus 0.5em minus 0.4em\relax Basic books, 2013.

\bibitem{okamura2020adaptive}
K.~Okamura and S.~Yamada, ``Adaptive trust calibration for human-ai
  collaboration,'' \emph{Plos one}, vol.~15, no.~2, p. e0229132, 2020.

\bibitem{yue2023impact}
B.~Yue and H.~Li, ``The impact of human-ai collaboration types on consumer
  evaluation and usage intention: a perspective of responsibility
  attribution,'' \emph{Frontiers in psychology}, vol.~14, p. 1277861, 2023.

\bibitem{dellermann2021future}
D.~Dellermann, A.~Calma, N.~Lipusch, T.~Weber, S.~Weigel, and P.~Ebel, ``The
  future of human-ai collaboration: a taxonomy of design knowledge for hybrid
  intelligence systems,'' \emph{arXiv preprint arXiv:2105.03354}, 2021.

\bibitem{seeber2020machines}
I.~Seeber, E.~Bittner, R.~O. Briggs, T.~De~Vreede, G.-J. De~Vreede, A.~Elkins,
  R.~Maier, A.~B. Merz, S.~Oeste-Rei{\ss}, N.~Randrup \emph{et~al.}, ``Machines
  as teammates: A research agenda on ai in team collaboration,''
  \emph{Information \& management}, vol.~57, no.~2, p. 103174, 2020.

\bibitem{crandall2018ishowo}
\BIBentryALTinterwordspacing
J.~W. Crandall, M.~Oudah, Tennom, F.~Ishowo-Oloko, S.~Abdallah, J.-F. Bonnefon,
  M.~Cebrian, A.~Shariff, M.~A. Goodrich, and I.~Rahwan, ``Cooperating with
  machines,'' \emph{Nature Communications}, vol.~9, no.~1, Jan. 2018. [Online].
  Available: \url{http://dx.doi.org/10.1038/s41467-017-02597-8}
\BIBentrySTDinterwordspacing

\bibitem{amershi2019guidelines}
S.~Amershi, D.~Weld, M.~Vorvoreanu, A.~Fourney, B.~Nushi, P.~Collisson, J.~Suh,
  S.~Iqbal, P.~N. Bennett, K.~Inkpen \emph{et~al.}, ``Guidelines for human-ai
  interaction,'' in \emph{Proceedings of the 2019 chi conference on human
  factors in computing systems}, 2019, pp. 1--13.

\bibitem{schwalbe2024comprehensive}
G.~Schwalbe and B.~Finzel, ``A comprehensive taxonomy for explainable
  artificial intelligence: a systematic survey of surveys on methods and
  concepts,'' \emph{Data Mining and Knowledge Discovery}, vol.~38, no.~5, pp.
  3043--3101, 2024.

\bibitem{dubey2020haco}
A.~Dubey, K.~Abhinav, S.~Jain, V.~Arora, and A.~Puttaveerana, ``Haco: a
  framework for developing human-ai teaming,'' in \emph{Proceedings of the 13th
  Innovations in Software Engineering Conference on Formerly known as India
  Software Engineering Conference}, 2020, pp. 1--9.

\bibitem{eigner2024determinants}
E.~Eigner and T.~H{\"a}ndler, ``Determinants of llm-assisted decision-making,''
  \emph{arXiv preprint arXiv:2402.17385}, 2024.

\bibitem{sutthithatip2021explainable}
S.~Sutthithatip, S.~Perinpanayagam, S.~Aslam, and A.~Wileman, ``Explainable ai
  in aerospace for enhanced system performance,'' in \emph{2021 IEEE/AIAA 40th
  Digital Avionics Systems Conference (DASC)}.\hskip 1em plus 0.5em minus
  0.4em\relax IEEE, 2021, pp. 1--7.

\bibitem{how2020artificial}
M.-L. How, S.-M. Cheah, Y.-J. Chan, A.~C. Khor, and E.~M.~P. Say, ``Artificial
  intelligence-enhanced decision support for informing global sustainable
  development: A human-centric ai-thinking approach,'' \emph{Information},
  vol.~11, no.~1, p.~39, 2020.

\bibitem{cesta2016towards}
A.~Cesta, A.~Orlandini, G.~Bernardi, and A.~Umbrico, ``Towards a planning-based
  framework for symbiotic human-robot collaboration,'' in \emph{2016 IEEE 21st
  international conference on emerging technologies and factory automation
  (ETFA)}.\hskip 1em plus 0.5em minus 0.4em\relax IEEE, 2016, pp. 1--8.

\bibitem{jokinen2019boundary}
K.~Jokinen and K.~Watanabe, ``Boundary-crossing robots: Societal impact of
  interactions with socially capable autonomous agents,'' in \emph{Social
  Robotics: 11th International Conference, ICSR 2019, Madrid, Spain, November
  26--29, 2019, Proceedings 11}.\hskip 1em plus 0.5em minus 0.4em\relax
  Springer, 2019, pp. 3--13.

\bibitem{mahmud2023study}
B.~Mahmud, G.~Hong, and B.~Fong, ``A study of human--ai symbiosis for creative
  work: Recent developments and future directions in deep learning,'' \emph{ACM
  Transactions on Multimedia Computing, Communications and Applications},
  vol.~20, no.~2, pp. 1--21, 2023.

\bibitem{sowa2021cobots}
K.~Sowa, A.~Przegalinska, and L.~Ciechanowski, ``Cobots in knowledge work:
  Human--ai collaboration in managerial professions,'' \emph{Journal of
  Business Research}, vol. 125, pp. 135--142, 2021.

\bibitem{alves2024cost}
J.~V. Alves, D.~Leit{\~a}o, S.~Jesus, M.~O. Sampaio, J.~Li{\'e}bana,
  P.~Saleiro, M.~A. Figueiredo, and P.~Bizarro, ``Cost-sensitive learning to
  defer to multiple experts with workload constraints,'' \emph{arXiv preprint
  arXiv:2403.06906}, 2024.

\bibitem{hemmer2022forming}
P.~Hemmer, S.~Schellhammer, M.~V{\"o}ssing, J.~Jakubik, and G.~Satzger,
  ``Forming effective human-ai teams: building machine learning models that
  complement the capabilities of multiple experts,'' \emph{arXiv preprint
  arXiv:2206.07948}, 2022.

\bibitem{liu2021role}
J.~Liu and Y.~Zhao, ``Role-oriented task allocation in human-machine
  collaboration system,'' in \emph{2021 IEEE 4th International Conference on
  Information Systems and Computer Aided Education (ICISCAE)}.\hskip 1em plus
  0.5em minus 0.4em\relax IEEE, 2021, pp. 243--248.

\bibitem{wang2020human}
D.~Wang, E.~Churchill, P.~Maes, X.~Fan, B.~Shneiderman, Y.~Shi, and Q.~Wang,
  ``From human-human collaboration to human-ai collaboration: Designing ai
  systems that can work together with people,'' in \emph{Extended abstracts of
  the 2020 CHI conference on human factors in computing systems}, 2020, pp.
  1--6.

\bibitem{vollmuth2023artificial}
P.~Vollmuth, M.~Foltyn, R.~Y. Huang, N.~Galldiks, J.~Petersen, F.~Isensee,
  M.~J. van~den Bent, F.~Barkhof, J.~E. Park, Y.~W. Park \emph{et~al.},
  ``Artificial intelligence (ai)-based decision support improves
  reproducibility of tumor response assessment in neuro-oncology: an
  international multi-reader study,'' \emph{Neuro-oncology}, vol.~25, no.~3,
  pp. 533--543, 2023.

\bibitem{van2022does}
K.~G. Van~Leeuwen, M.~de~Rooij, S.~Schalekamp, B.~van Ginneken, and M.~J.
  Rutten, ``How does artificial intelligence in radiology improve efficiency
  and health outcomes?'' \emph{Pediatric Radiology}, pp. 1--7, 2022.

\bibitem{timmons2023call}
A.~C. Timmons, J.~B. Duong, N.~Simo~Fiallo, T.~Lee, H.~P.~Q. Vo, M.~W. Ahle,
  J.~S. Comer, L.~C. Brewer, S.~L. Frazier, and T.~Chaspari, ``A call to action
  on assessing and mitigating bias in artificial intelligence applications for
  mental health,'' \emph{Perspectives on Psychological Science}, vol.~18,
  no.~5, pp. 1062--1096, 2023.

\bibitem{farivc2024early}
N.~Fari{\v{c}}, S.~Hinder, R.~Williams, R.~Ramaesh, M.~O. Bernabeu, E.~van
  Beek, and K.~Cresswell, ``Early experiences of integrating an artificial
  intelligence-based diagnostic decision support system into radiology
  settings: a qualitative study,'' \emph{Journal of the American Medical
  Informatics Association}, vol.~31, no.~1, pp. 24--34, 2024.

\bibitem{calisto2022breastscreening}
F.~M. Calisto, C.~Santiago, N.~Nunes, and J.~C. Nascimento,
  ``Breastscreening-ai: Evaluating medical intelligent agents for human-ai
  interactions,'' \emph{Artificial Intelligence in Medicine}, vol. 127, p.
  102285, 2022.

\bibitem{rezwana2023designing}
J.~Rezwana and M.~L. Maher, ``Designing creative ai partners with cofi: A
  framework for modeling interaction in human-ai co-creative systems,''
  \emph{ACM Transactions on Computer-Human Interaction}, vol.~30, no.~5, pp.
  1--28, 2023.

\bibitem{siu2021evaluation}
H.~C. Siu, J.~Pe{\~n}a, E.~Chen, Y.~Zhou, V.~Lopez, K.~Palko, K.~Chang, and
  R.~Allen, ``Evaluation of human-ai teams for learned and rule-based agents in
  hanabi,'' \emph{Advances in Neural Information Processing Systems}, vol.~34,
  pp. 16\,183--16\,195, 2021.

\bibitem{reddy2021evaluation}
S.~Reddy, W.~Rogers, V.-P. Makinen, E.~Coiera, P.~Brown, M.~Wenzel, E.~Weicken,
  S.~Ansari, P.~Mathur, A.~Casey \emph{et~al.}, ``Evaluation framework to guide
  implementation of ai systems into healthcare settings,'' \emph{BMJ health \&
  care informatics}, vol.~28, no.~1, 2021.

\bibitem{sharma2023human}
A.~Sharma, I.~W. Lin, A.~S. Miner, D.~C. Atkins, and T.~Althoff, ``Human--ai
  collaboration enables more empathic conversations in text-based peer-to-peer
  mental health support,'' \emph{Nature Machine Intelligence}, vol.~5, no.~1,
  pp. 46--57, 2023.

\bibitem{yik2024neurobench}
J.~Yik, K.~V. den Berghe, D.~den Blanken, Y.~Bouhadjar, M.~Fabre, P.~Hueber,
  D.~Kleyko, and et~al., ``Neurobench: A framework for benchmarking
  neuromorphic computing algorithms and systems,'' 2024.

\bibitem{rastogi2023supporting}
C.~Rastogi, M.~Tulio~Ribeiro, N.~King, H.~Nori, and S.~Amershi, ``Supporting
  human-ai collaboration in auditing llms with llms,'' in \emph{Proceedings of
  the 2023 AAAI/ACM Conference on AI, Ethics, and Society}, 2023, pp. 913--926.

\bibitem{zhao_chatspot_2023}
\BIBentryALTinterwordspacing
L.~Zhao, E.~Yu, Z.~Ge, J.~Yang, H.~Wei, H.~Zhou, J.~Sun, Y.~Peng, R.~Dong,
  C.~Han, and X.~Zhang, ``{ChatSpot}: {Bootstrapping} {Multimodal} {LLMs} via
  {Precise} { Referring} {Instruction} {Tuning},'' Jul. 2023, arXiv:2307.09474
  [cs] version: 1. [Online]. Available: \url{http://arxiv.org/abs/2307.09474}
\BIBentrySTDinterwordspacing

\bibitem{aleid2023artificial}
A.~Aleid, K.~Alhussaini, R.~Alanazi, M.~Altwaimi, O.~Altwijri, and A.~S. Saad,
  ``Artificial intelligence approach for early detection of brain tumors using
  mri images,'' \emph{Applied Sciences}, vol.~13, no.~6, p. 3808, 2023.

\bibitem{al2023ai}
A.~Al-Fatlawi, A.~A.~T. Al-Khazaali, S.~H. Hasan \emph{et~al.}, ``Ai-based
  model for fraud detection in bank systems,'' \emph{Fusion: Practice and
  Applications}, vol.~14, no.~1, pp. 19--9, 2023.

\bibitem{yang2022optimal}
M.~Yang, M.~Carroll, and A.~Dragan, ``Optimal behavior prior: Data-efficient
  human models for improved human-ai collaboration,'' \emph{arXiv preprint
  arXiv:2211.01602}, 2022.

\bibitem{aher2023using}
G.~V. Aher, R.~I. Arriaga, and A.~T. Kalai, ``Using large language models to
  simulate multiple humans and replicate human subject studies,'' in
  \emph{International Conference on Machine Learning}.\hskip 1em plus 0.5em
  minus 0.4em\relax PMLR, 2023, pp. 337--371.

\bibitem{verheijden2023collaborative}
M.~P. Verheijden and M.~Funk, ``Collaborative diffusion: Boosting designerly
  co-creation with generative ai,'' in \emph{Extended Abstracts of the 2023 CHI
  Conference on Human Factors in Computing Systems}, 2023, pp. 1--8.

\bibitem{hauptman2023adapt}
A.~I. Hauptman, B.~G. Schelble, N.~J. McNeese, and K.~C. Madathil, ``Adapt and
  overcome: Perceptions of adaptive autonomous agents for human-ai teaming,''
  \emph{Computers in Human Behavior}, vol. 138, p. 107451, 2023.

\bibitem{sankaran2022modeling}
G.~Sankaran, M.~A. Palomino, M.~Knahl, and G.~Siestrup, ``A modeling approach
  for measuring the performance of a human-ai collaborative process,''
  \emph{Applied Sciences}, vol.~12, no.~22, p. 11642, 2022.

\bibitem{massaro_multi-level_2022}
\BIBentryALTinterwordspacing
A.~Massaro, ``\BIBforeignlanguage{en}{Multi-{Level} {Decision} {Support}
  {System} in {Production} and { Safety} {Management}},''
  \emph{\BIBforeignlanguage{en}{Knowledge}}, vol.~2, no.~4, pp. 682--701, Dec.
  2022. [Online]. Available: \url{https://www.mdpi.com/2673-9585/2/4/39}
\BIBentrySTDinterwordspacing

\bibitem{oravec2023artificial}
J.~A. Oravec, ``Artificial intelligence implications for academic cheating:
  Expanding the dimensions of responsible human-ai collaboration with
  chatgpt,'' \emph{Journal of Interactive Learning Research}, vol.~34, no.~2,
  pp. 213--237, 2023.

\bibitem{el-zanfaly_sand_2022}
\BIBentryALTinterwordspacing
D.~EL-Zanfaly, Y.~Huang, and Y.~Dong, ``\BIBforeignlanguage{en}{Sand
  {Playground}: {Designing} {Human}-{AI} physical {Interface} for {Co}-creation
  in {Motion}},'' in \emph{\BIBforeignlanguage{en}{Creativity and
  {Cognition}}}.\hskip 1em plus 0.5em minus 0.4em\relax Venice Italy: ACM, Jun.
  2022, pp. 49--55. [Online]. Available:
  \url{https://dl.acm.org/doi/10.1145/3527927.3532791}
\BIBentrySTDinterwordspacing

\bibitem{arias-rosales_perceived_2022}
\BIBentryALTinterwordspacing
A.~Arias-Rosales, ``\BIBforeignlanguage{en}{The perceived value of human-{AI}
  collaboration in early shape exploration: {An} exploratory assessment},''
  \emph{\BIBforeignlanguage{en}{PLOS ONE}}, vol.~17, no.~9, p. e0274496, Sep.
  2022. [Online]. Available:
  \url{https://dx.plos.org/10.1371/journal.pone.0274496}
\BIBentrySTDinterwordspacing

\bibitem{dziorny_clinical_2022}
\BIBentryALTinterwordspacing
A.~C. Dziorny, J.~A. Heneghan, M.~A. Bhat, D.~J. Karavite, L.~N. Sanchez-Pinto,
  J.~McArthur, and N.~Muthu, ``Clinical decision support in the {PICU}:
  {Implications} for design and evaluation,'' \emph{Pediatric Critical Care
  Medicine}, vol.~23, no.~8, pp. e392--e396, 2022, publisher: LWW. [Online].
  Available:
  \url{https://journals.lww.com/pccmjournal/Fulltext/2022/08000/Clinical_Decision_Support_in_the_PICU_.19.aspx}
\BIBentrySTDinterwordspacing

\bibitem{vossing2022designing}
M.~V{\"o}ssing, N.~K{\"u}hl, M.~Lind, and G.~Satzger, ``Designing transparency
  for effective human-ai collaboration,'' \emph{Information Systems Frontiers},
  vol.~24, no.~3, pp. 877--895, 2022.

\bibitem{chakravorti2022artificial}
T.~Chakravorti, V.~Singh, S.~Rajtmajer, M.~McLaughlin, R.~Fraleigh, C.~Griffin,
  A.~Kwasnica, D.~Pennock, and C.~L. Giles, ``Artificial prediction markets
  present a novel opportunity for human-ai collaboration,'' \emph{arXiv
  preprint arXiv:2211.16590}, 2022.

\bibitem{huang2022framework}
M.-H. Huang and R.~T. Rust, ``A framework for collaborative artificial
  intelligence in marketing,'' \emph{Journal of Retailing}, vol.~98, no.~2, pp.
  209--223, 2022.

\bibitem{sachan2024human}
S.~Sachan, F.~Almaghrabi, J.-B. Yang, and D.-L. Xu, ``Human-ai collaboration to
  mitigate decision noise in financial underwriting: A study on fintech
  innovation in a lending firm,'' \emph{International Review of Financial
  Analysis}, p. 103149, 2024.

\bibitem{basu2021human}
S.~Basu, A.~Garimella, W.~Han, and A.~Dennis, ``Human decision making in ai
  augmented systems: Evidence from the initial coin offering market,'' 2021.

\bibitem{dikmen2022effects}
M.~Dikmen and C.~Burns, ``The effects of domain knowledge on trust in
  explainable ai and task performance: A case of peer-to-peer lending,''
  \emph{International Journal of Human-Computer Studies}, vol. 162, p. 102792,
  2022.

\bibitem{ma2024towards}
S.~Ma, Q.~Chen, X.~Wang, C.~Zheng, Z.~Peng, M.~Yin, and X.~Ma, ``Towards
  human-ai deliberation: Design and evaluation of llm-empowered deliberative ai
  for ai-assisted decision-making,'' \emph{arXiv preprint arXiv:2403.16812},
  2024.

\bibitem{alon2023human}
S.~Alon-Barkat and M.~Busuioc, ``Human--ai interactions in public sector
  decision making:“automation bias” and “selective adherence” to
  algorithmic advice,'' \emph{Journal of Public Administration Research and
  Theory}, vol.~33, no.~1, pp. 153--169, 2023.

\bibitem{nikolopoulou2024generative}
K.~Nikolopoulou, ``Generative artificial intelligence in higher education:
  Exploring ways of harnessing pedagogical practices with the assistance of
  chatgpt,'' \emph{International Journal of Changes in Education}, 2024.

\bibitem{fischer2023future}
F.~Fischer, ``Future collaboration between humans and ai: “i strive to make
  you feel good,” says my ai colleague in 2030,'' in \emph{Work and AI 2030:
  Challenges and Strategies for Tomorrow's Work}.\hskip 1em plus 0.5em minus
  0.4em\relax Springer, 2023, pp. 21--28.

\bibitem{nasir2024ethical}
S.~Nasir, R.~A. Khan, and S.~Bai, ``Ethical framework for harnessing the power
  of ai in healthcare and beyond,'' \emph{IEEE Access}, vol.~12, pp.
  31\,014--31\,035, 2024.

\bibitem{lase2023human}
E.~M. Lase and F.~Nkosi, ``Human-centric ai: Understanding and enhancing
  collaboration between humans and intelligent systems,'' \emph{Algorithm
  Asynchronous}, vol.~1, no.~1, pp. 33--40, 2023.

\bibitem{bojic2023hierarchical}
I.~Bojic, J.~Chen, S.~Y. Chang, Q.~C. Ong, S.~Joty, and J.~Car, ``Hierarchical
  evaluation framework: Best practices for human evaluation,'' \emph{arXiv
  preprint arXiv:2310.01917}, 2023.

\bibitem{saha2023human}
G.~C. Saha, S.~Kumar, A.~Kumar, H.~Saha, T.~Lakshmi, and N.~Bhat, ``Human-ai
  collaboration: Exploring interfaces for interactive machine learning,''
  \emph{Tuijin Jishu/Journal of Propulsion Technology}, vol.~44, no.~2, p.
  2023, 2023.

\bibitem{vossing_designing_2022}
\BIBentryALTinterwordspacing
M.~Vössing, N.~Kühl, M.~Lind, and G.~Satzger,
  ``\BIBforeignlanguage{en}{Designing {Transparency} for {Effective}
  {Human}-{AI} { Collaboration}},'' \emph{\BIBforeignlanguage{en}{Information
  Systems Frontiers}}, vol.~24, no.~3, pp. 877--895, Jun. 2022. [Online].
  Available: \url{https://doi.org/10.1007/s10796-022-10284-3}
\BIBentrySTDinterwordspacing

\bibitem{fabri2023disentangling}
L.~Fabri, B.~H{\"a}ckel, A.~M. Oberl{\"a}nder, M.~Rieg, and A.~Stohr,
  ``Disentangling human-ai hybrids,'' \emph{Business \& information systems
  engineering}, pp. 1--19, 2023.

\bibitem{wu2021ai}
Z.~Wu, D.~Ji, K.~Yu, X.~Zeng, D.~Wu, and M.~Shidujaman, ``Ai creativity and the
  human-ai co-creation model,'' in \emph{Human-Computer Interaction. Theory,
  Methods and Tools: Thematic Area, HCI 2021, Held as Part of the 23rd HCI
  International Conference, HCII 2021, Virtual Event, July 24--29, 2021,
  Proceedings, Part I 23}.\hskip 1em plus 0.5em minus 0.4em\relax Springer,
  2021, pp. 171--190.

\bibitem{li_human-ai_2022}
\BIBentryALTinterwordspacing
J.~Li, J.~Huang, J.~Liu, and T.~Zheng, ``\BIBforeignlanguage{en}{Human-{AI}
  cooperation: {Modes} and their effects on attitudes},''
  \emph{\BIBforeignlanguage{en}{Telematics and Informatics}}, vol.~73, p.
  101862, Sep. 2022. [Online]. Available:
  \url{https://linkinghub.elsevier.com/retrieve/pii/S0736585322000958}
\BIBentrySTDinterwordspacing

\bibitem{abedin2022designing}
B.~Abedin, C.~Meske, I.~Junglas, F.~Rabhi, and H.~R. Motahari-Nezhad,
  ``Designing and managing human-ai interactions,'' \emph{Information Systems
  Frontiers}, vol.~24, no.~3, pp. 691--697, 2022.

\bibitem{xiong_challenges_2022}
\BIBentryALTinterwordspacing
W.~Xiong, H.~Fan, L.~Ma, and C.~Wang, ``\BIBforeignlanguage{en}{Challenges of
  human—machine collaboration in risky decision-making},''
  \emph{\BIBforeignlanguage{en}{Frontiers of Engineering Management}}, vol.~9,
  no.~1, pp. 89--103, Mar. 2022. [Online]. Available:
  \url{https://link.springer.com/10.1007/s42524-021-0182-0}
\BIBentrySTDinterwordspacing

\bibitem{cakmak2014eliciting}
M.~Cakmak and A.~L. Thomaz, ``Eliciting good teaching from humans for machine
  learners,'' \emph{Artificial Intelligence}, vol. 217, pp. 198--215, 2014.

\bibitem{powers2020evaluation}
D.~M. Powers, ``Evaluation: from precision, recall and f-measure to roc,
  informedness, markedness and correlation,'' \emph{arXiv preprint
  arXiv:2010.16061}, 2020.

\bibitem{kerzner2017project}
H.~Kerzner, \emph{Project management: a systems approach to planning,
  scheduling, and controlling}.\hskip 1em plus 0.5em minus 0.4em\relax John
  Wiley \& Sons, 2017.

\bibitem{hinsen2022can}
S.~Hinsen, P.~Hofmann, J.~J{\"o}hnk, and N.~Urbach, ``How can organizations
  design purposeful human-ai interactions: a practical perspective from
  existing use cases and interviews,'' 2022.

\bibitem{amershi2014power}
S.~Amershi, M.~Cakmak, W.~B. Knox, and T.~Kulesza, ``Power to the people: The
  role of humans in interactive machine learning,'' \emph{Ai Magazine},
  vol.~35, no.~4, pp. 105--120, 2014.

\bibitem{dautenhahn2007socially}
K.~Dautenhahn, ``Socially intelligent robots: dimensions of human--robot
  interaction,'' \emph{Philosophical transactions of the royal society B:
  Biological sciences}, vol. 362, no. 1480, pp. 679--704, 2007.

\bibitem{wenskovitch2020interactive}
J.~Wenskovitch and C.~North, ``Interactive artificial intelligence: designing
  for the" two black boxes" problem,'' \emph{Computer}, vol.~53, no.~8, pp.
  29--39, 2020.

\bibitem{el2022biases}
M.~El-Assady and C.~Moruzzi, ``Which biases and reasoning pitfalls do
  explanations trigger? decomposing communication processes in human--ai
  interaction,'' \emph{IEEE Computer Graphics and Applications}, vol.~42,
  no.~6, pp. 11--23, 2022.

\bibitem{papenmeier2022s}
A.~Papenmeier, D.~Kern, G.~Englebienne, and C.~Seifert, ``It’s complicated:
  The relationship between user trust, model accuracy and explanations in ai,''
  \emph{ACM Transactions on Computer-Human Interaction (TOCHI)}, vol.~29,
  no.~4, pp. 1--33, 2022.

\bibitem{magrabi2010analysis}
F.~Magrabi, M.-S. Ong, W.~Runciman, and E.~Coiera, ``An analysis of
  computer-related patient safety incidents to inform the development of a
  classification,'' \emph{Journal of the American Medical Informatics
  Association}, vol.~17, no.~6, pp. 663--670, 2010.

\bibitem{zhang2020effect}
Y.~Zhang, Q.~V. Liao, and R.~K. Bellamy, ``Effect of confidence and explanation
  on accuracy and trust calibration in ai-assisted decision making,'' in
  \emph{Proceedings of the 2020 conference on fairness, accountability, and
  transparency}, 2020, pp. 295--305.

\bibitem{lopes2023towards}
R.~Lopes, D.~Raposo, and S.~Sargento, ``Towards time sensitive networking on
  smart cities: Techniques, challenges, and solutions,'' \emph{arXiv preprint
  arXiv:2312.03635}, 2023.

\bibitem{brady2024developing}
A.~P. Brady, B.~Allen, J.~Chong, E.~Kotter, N.~Kottler, J.~Mongan,
  L.~Oakden-Rayner, D.~P. Dos~Santos, A.~Tang, C.~Wald \emph{et~al.},
  ``Developing, purchasing, implementing and monitoring ai tools in radiology:
  practical considerations. a multi-society statement from the acr, car, esr,
  ranzcr \& rsna,'' \emph{Insights into Imaging}, vol.~15, no.~1, p.~16, 2024.

\bibitem{zahedi_human-ai_2021}
\BIBentryALTinterwordspacing
Z.~Zahedi and S.~Kambhampati, ``\BIBforeignlanguage{en}{Human-{AI} {Symbiosis}:
  {A} {Survey} of {Current} {Approaches}},'' Mar. 2021, arXiv:2103.09990 [cs].
  [Online]. Available: \url{http://arxiv.org/abs/2103.09990}
\BIBentrySTDinterwordspacing

\bibitem{schneider2023assessing}
M.~F. Schneider, M.~E. Miller, and J.~McGuirl, ``Assessing quality goal
  rankings as a method for communicating operator intent,'' \emph{Journal of
  Cognitive Engineering and Decision Making}, vol.~17, no.~1, pp. 26--48, 2023.

\bibitem{mikalef2021artificial}
P.~Mikalef and M.~Gupta, ``Artificial intelligence capability:
  Conceptualization, measurement calibration, and empirical study on its impact
  on organizational creativity and firm performance,'' \emph{Information \&
  Management}, vol.~58, no.~3, p. 103434, 2021.

\bibitem{shafiq2022student}
D.~A. Shafiq, M.~Marjani, R.~A.~A. Habeeb, and D.~Asirvatham, ``Student
  retention using educational data mining and predictive analytics: a
  systematic literature review,'' \emph{IEEE Access}, 2022.

\bibitem{yang2022user}
R.~Yang and S.~Wibowo, ``User trust in artificial intelligence: A comprehensive
  conceptual framework,'' \emph{Electronic Markets}, vol.~32, no.~4, pp.
  2053--2077, 2022.

\bibitem{murugesan2023study}
U.~Murugesan, P.~Subramanian, S.~Srivastava, and A.~Dwivedi, ``A study of
  artificial intelligence impacts on human resource digitalization in industry
  4.0,'' \emph{Decision Analytics Journal}, p. 100249, 2023.

\bibitem{gokhale2022generalized}
T.~Gokhale, S.~Mishra, M.~Luo, B.~S. Sachdeva, and C.~Baral, ``Generalized but
  not robust? comparing the effects of data modification methods on
  out-of-domain generalization and adversarial robustness,'' \emph{arXiv
  preprint arXiv:2203.07653}, 2022.

\bibitem{kuo2000annotated}
W.~Kuo and V.~R. Prasad, ``An annotated overview of system-reliability
  optimization,'' \emph{IEEE Transactions on reliability}, vol.~49, no.~2, pp.
  176--187, 2000.

\bibitem{fiebrink2011human}
R.~Fiebrink, P.~R. Cook, and D.~Trueman, ``Human model evaluation in
  interactive supervised learning,'' pp. 147--156, 2011.

\bibitem{hoc2000human}
J.-M. Hoc, ``From human--machine interaction to human--machine cooperation,''
  \emph{Ergonomics}, vol.~43, no.~7, pp. 833--843, 2000.

\bibitem{dehghani2024trustworthy}
F.~Dehghani, M.~Dibaji, F.~Anzum, L.~Dey, A.~Basdemir, S.~Bayat, J.-C. Boucher,
  S.~Drew, S.~E. Eaton, R.~Frayne \emph{et~al.}, ``Trustworthy and responsible
  ai for human-centric autonomous decision-making systems,'' \emph{arXiv
  preprint arXiv:2408.15550}, 2024.

\bibitem{hein2024acceptance}
I.~Hein, J.~Cecil, and E.~Lermer, ``Acceptance and motivational effect of
  ai-driven feedback in the workplace: an experimental study with direct
  replication,'' \emph{Frontiers in Organizational Psychology}, vol.~2, p.
  1468907, 2024.

\bibitem{inkpen2023advancing}
K.~Inkpen, S.~Chappidi, K.~Mallari, B.~Nushi, D.~Ramesh, P.~Michelucci,
  V.~Mandava, L.~H. Vep{\v{r}}ek, and G.~Quinn, ``Advancing human-ai
  complementarity: The impact of user expertise and algorithmic tuning on joint
  decision making,'' \emph{ACM Transactions on Computer-Human Interaction},
  vol.~30, no.~5, pp. 1--29, 2023.

\bibitem{lee1994trust}
J.~D. Lee and N.~Moray, ``Trust, self-confidence, and operators' adaptation to
  automation,'' \emph{International journal of human-computer studies},
  vol.~40, no.~1, pp. 153--184, 1994.

\bibitem{barocas2023fairness}
S.~Barocas, M.~Hardt, and A.~Narayanan, \emph{Fairness and machine learning:
  Limitations and opportunities}.\hskip 1em plus 0.5em minus 0.4em\relax MIT
  press, 2023.

\bibitem{holstein2019improving}
K.~Holstein, J.~Wortman~Vaughan, H.~Daum{\'e}~III, M.~Dudik, and H.~Wallach,
  ``Improving fairness in machine learning systems: What do industry
  practitioners need?'' in \emph{Proceedings of the 2019 CHI conference on
  human factors in computing systems}, 2019, pp. 1--16.

\bibitem{horvitz1999principles}
E.~Horvitz, ``Principles of mixed-initiative user interfaces,'' in
  \emph{Proceedings of the SIGCHI conference on Human Factors in Computing
  Systems}, 1999, pp. 159--166.

\bibitem{tsarouchi2017human}
P.~Tsarouchi, A.-S. Matthaiakis, S.~Makris, and G.~Chryssolouris, ``On a
  human-robot collaboration in an assembly cell,'' \emph{International Journal
  of Computer Integrated Manufacturing}, vol.~30, no.~6, pp. 580--589, 2017.

\bibitem{hemmer2022factors}
P.~Hemmer, M.~Schemmer, L.~Riefle, N.~Rosellen, M.~V{\"o}ssing, and
  N.~K{\"u}hl, ``Factors that influence the adoption of human-ai collaboration
  in clinical decision-making,'' \emph{arXiv preprint arXiv:2204.09082}, 2022.

\bibitem{yin2021group}
X.~Yin, J.~Huang, W.~He, W.~Guo, H.~Yu, and L.~Cui, ``Group task allocation
  approach for heterogeneous software crowdsourcing tasks,'' \emph{Peer-to-Peer
  Networking and Applications}, vol.~14, pp. 1736--1747, 2021.

\bibitem{fischer1995rethinking}
G.~Fischer, ``Rethinking and reinventing artificial intelligence from the
  perspective of human-centered computational artifacts,'' in \emph{Brazilian
  Symposium on Artificial Intelligence}.\hskip 1em plus 0.5em minus 0.4em\relax
  Springer, 1995, pp. 1--11.

\bibitem{kamar2012combining}
E.~Kamar, S.~Hacker, and E.~Horvitz, ``Combining human and machine intelligence
  in large-scale crowdsourcing.'' in \emph{AAMAS}, vol.~12, 2012, pp. 467--474.

\bibitem{rabby2020modeling}
M.~K.~M. Rabby, M.~A. Khan, A.~Karimoddini, and S.~X. Jiang, ``Modeling of
  trust within a human-robot collaboration framework,'' in \emph{2020 IEEE
  International Conference on Systems, Man, and Cybernetics (SMC)}.\hskip 1em
  plus 0.5em minus 0.4em\relax IEEE, 2020, pp. 4267--4272.

\bibitem{mohammadi2020mixed}
F.~Mohammadi~Amin, M.~Rezayati, H.~W. van~de Venn, and H.~Karimpour, ``A
  mixed-perception approach for safe human--robot collaboration in industrial
  automation,'' \emph{Sensors}, vol.~20, no.~21, p. 6347, 2020.

\bibitem{hou2021enabling}
M.~Hou, ``Enabling trust in autonomous human-machine teaming,'' in \emph{2021
  IEEE International Conference on Autonomous Systems (ICAS)}.\hskip 1em plus
  0.5em minus 0.4em\relax IEEE, 2021, pp. 1--1.

\bibitem{liu2019comparison}
X.~Liu, L.~Faes, A.~U. Kale, S.~K. Wagner, D.~J. Fu, A.~Bruynseels,
  T.~Mahendiran, G.~Moraes, M.~Shamdas, C.~Kern \emph{et~al.}, ``A comparison
  of deep learning performance against health-care professionals in detecting
  diseases from medical imaging: a systematic review and meta-analysis,''
  \emph{The lancet digital health}, vol.~1, no.~6, pp. e271--e297, 2019.

\bibitem{zheng2023overview}
D.~Zheng, X.~He, and J.~Jing, ``Overview of artificial intelligence in breast
  cancer medical imaging,'' \emph{Journal of Clinical Medicine}, vol.~12,
  no.~2, p. 419, 2023.

\bibitem{kaboudan2023ai}
A.~Kaboudan and W.~Salah~Eldin, ``Ai-driven medical imaging platform:
  Advancements in image analysis and healthcare diagnosis,'' \emph{Journal of
  the ACS Advances in Computer Science}, vol.~14, no.~1, 2023.

\bibitem{dziorny2022clinical}
A.~C. Dziorny, J.~A. Heneghan, M.~A. Bhat, D.~J. Karavite, L.~N. Sanchez-Pinto,
  J.~McArthur, N.~Muthu \emph{et~al.}, ``Clinical decision support in the picu:
  Implications for design and evaluation,'' \emph{Pediatric Critical Care
  Medicine}, vol.~23, no.~8, pp. e392--e396, 2022.

\bibitem{atadoga2024ai}
A.~Atadoga, O.~C. Obi, S.~Onwusinkwue, S.~O. Dawodu, F.~Osasona, A.~I.
  Daraojimba \emph{et~al.}, ``Ai's evolving impact in us banking: An insightful
  review,'' \emph{International Journal of Science and Research Archive},
  vol.~11, no.~1, pp. 904--922, 2024.

\bibitem{jain2023role}
R.~Jain, ``Role of artificial intelligence in banking and finance,''
  \emph{Journal of Management and Science}, vol.~13, no.~3, pp. 1--4, 2023.

\bibitem{ali2022financial}
A.~Ali, S.~Abd~Razak, S.~H. Othman, T.~A.~E. Eisa, A.~Al-Dhaqm, M.~Nasser,
  T.~Elhassan, H.~Elshafie, and A.~Saif, ``Financial fraud detection based on
  machine learning: a systematic literature review,'' \emph{Applied Sciences},
  vol.~12, no.~19, p. 9637, 2022.

\bibitem{ifenthaler2023reciprocal}
D.~Ifenthaler and C.~Schumacher, ``Reciprocal issues of artificial and human
  intelligence in education,'' pp. 1--6, 2023.

\bibitem{ji2023systematic}
H.~Ji, I.~Han, and Y.~Ko, ``A systematic review of conversational ai in
  language education: Focusing on the collaboration with human teachers,''
  \emph{Journal of Research on Technology in Education}, vol.~55, no.~1, pp.
  48--63, 2023.

\bibitem{fadiya2023development}
O.~Fadiya, ``Development of a student-centred manual using appreciative
  inquiry,'' \emph{International Journal of Quality and Service Sciences}, no.
  ahead-of-print, 2023.

\bibitem{kublik2022gpt}
S.~Kublik and S.~Saboo, \emph{GPT-3}.\hskip 1em plus 0.5em minus 0.4em\relax
  O'Reilly Media, Incorporated, 2022.

\bibitem{kalyan2023survey}
K.~S. Kalyan, ``A survey of gpt-3 family large language models including
  chatgpt and gpt-4,'' \emph{Natural Language Processing Journal}, p. 100048,
  2023.

\bibitem{vats2024survey}
V.~Vats, M.~B. Nizam, M.~Liu, Z.~Wang, R.~Ho, M.~S. Prasad, V.~Titterton, S.~V.
  Malreddy, R.~Aggarwal, Y.~Xu \emph{et~al.}, ``A survey on human-ai teaming
  with large pre-trained models,'' \emph{arXiv preprint arXiv:2403.04931},
  2024.

\bibitem{binz2023using}
M.~Binz and E.~Schulz, ``Using cognitive psychology to understand gpt-3,''
  \emph{Proceedings of the National Academy of Sciences}, vol. 120, no.~6, p.
  e2218523120, 2023.

\bibitem{webb2023emergent}
T.~Webb, K.~J. Holyoak, and H.~Lu, ``Emergent analogical reasoning in large
  language models,'' \emph{Nature Human Behaviour}, vol.~7, no.~9, pp.
  1526--1541, 2023.

\bibitem{douglas2023large}
M.~R. Douglas, ``Large language models,'' \emph{arXiv preprint
  arXiv:2307.05782}, 2023.

\bibitem{liu2023summary}
Y.~Liu, T.~Han, S.~Ma, J.~Zhang, Y.~Yang, J.~Tian, H.~He, A.~Li, M.~He, Z.~Liu
  \emph{et~al.}, ``Summary of chatgpt-related research and perspective towards
  the future of large language models,'' \emph{Meta-Radiology}, p. 100017,
  2023.

\bibitem{yin2023large}
Z.~Yin, Q.~Sun, Q.~Guo, J.~Wu, X.~Qiu, and X.~Huang, ``Do large language models
  know what they don't know?'' 2023.

\bibitem{dwivedi2023so}
Y.~K. Dwivedi, N.~Kshetri, L.~Hughes, E.~L. Slade, A.~Jeyaraj, A.~K. Kar, A.~M.
  Baabdullah, A.~Koohang, V.~Raghavan, M.~Ahuja \emph{et~al.}, ``“so what if
  chatgpt wrote it?” multidisciplinary perspectives on opportunities,
  challenges and implications of generative conversational ai for research,
  practice and policy,'' \emph{International Journal of Information
  Management}, vol.~71, p. 102642, 2023.

\bibitem{gmeiner2022team}
F.~Gmeiner, K.~Holstein, and N.~Martelaro, ``Team learning as a lens for
  designing human-ai co-creative systems,'' 2022.

\bibitem{jiang2023ai}
H.~H. Jiang, L.~Brown, J.~Cheng, M.~Khan, A.~Gupta, D.~Workman, A.~Hanna,
  J.~Flowers, and T.~Gebru, ``Ai art and its impact on artists,'' in
  \emph{Proceedings of the 2023 AAAI/ACM Conference on AI, Ethics, and
  Society}, 2023, pp. 363--374.

\bibitem{he2023searching}
V.~F. He, Y.~R. Shrestha, P.~Puranam, and E.~Miron-Spektor, ``Searching
  together: A theory of human-ai co-creativity,'' 2023.

\bibitem{epstein2023art}
Z.~Epstein, A.~Hertzmann, I.~of~Human~Creativity, M.~Akten, H.~Farid, J.~Fjeld,
  M.~R. Frank, M.~Groh, L.~Herman, N.~Leach \emph{et~al.}, ``Art and the
  science of generative ai,'' \emph{Science}, vol. 380, no. 6650, pp.
  1110--1111, 2023.

\bibitem{suh2021ai}
M.~Suh, E.~Youngblom, M.~Terry, and C.~J. Cai, ``Ai as social glue: uncovering
  the roles of deep generative ai during social music composition,'' in
  \emph{Proceedings of the 2021 CHI conference on human factors in computing
  systems}, 2021, pp. 1--11.

\end{thebibliography}
\newpage
\appendix

\section{The Unique Challenge of Evaluating Creative and Linguistic AI}\label{app:llm_genai}

Establishing comprehensive frameworks for assessing HAIC is vital, however certain applications pose distinct challenges due to their unique nature. Two prominent examples are
Large Language Models (LLMs) and Generative AI in the Arts, where AI's impact on human creativity, expression, and communication requires specialized evaluation methodologies.

While our framework provides a foundation, traditional metrics may be inadequate for these domains. Can quantitative measures fully capture artistic collaboration or AI’s influence on human language? How do we assess ethical concerns, bias, or AI-generated content’s societal impact? These challenges call for a nuanced evaluation approach. Although we don't address these challenges in this work, we briefly discuss them here for completeness.

\subsection{Large Language Models (LLMs)}
LLMs have transformed text generation, translation, and dialogue. The GPT-3~\cite{kublik2022gpt} family, including ChatGPT and GPT-4, demonstrates their capacity for complex NLP tasks~\cite{kalyan2023survey}. Evaluating HAIC with LLMs requires developing methodologies beyond traditional metrics~\cite{vats2024survey}.

Ensuring interpretability and fairness is vital. Binz \& Schulz (2022)\cite{binz2023using} show that even high-performing LLMs may use opaque reasoning patterns. Evaluation frameworks must assess how well LLMs explain their outputs and mitigate biases. Aher et al. (2022)\cite{aher2023using} highlight the importance of testing LLM adaptability in user interactions.

LLMs augment cognition, enhance creativity, and personalize learning. Webb et al. (2022)\cite{webb2023emergent} show their potential in analogical reasoning. Douglas (2023)\cite{douglas2023large} emphasizes their impact on ideation and artistic expression, while Liu et al. (2023)~\cite{liu2023summary} explore their role in education. Evaluating these aspects requires metrics for creativity, knowledge retention, and learner autonomy.

Longitudinal studies provide insights into HAIC with LLMs. Yin et al. (2023)\cite{yin2023large} suggest tracking how AI capabilities and user trust evolve over time. Dwivedi et al. (2023)~\cite{dwivedi2023so} address societal implications, underscoring the need for ethical AI use.

Developing robust LLM evaluation frameworks is essential. Addressing interpretability, bias, and interaction quality will help create responsible, effective HAIC systems.

\subsection{Generative AI in the Arts}

Generative AI is reshaping artistic expression across visual art, music, and poetry. AI moves beyond tools to co-creators, expanding artistic possibilities. Notable examples include AI-generated paintings, compositions, and interactive installations. Gmeiner et al. (2022)~\cite{gmeiner2022team} suggest leveraging team learning frameworks to enhance AI collaboration in creative fields.

Evaluating HAIC in generative AI requires adapting traditional art metrics—such as aesthetic quality, originality, and audience reception—to consider AI’s role in the creative process~\cite{jiang2023ai}. Verheijden \& Funk (2023)~\cite{verheijden2023collaborative} stress assessing AI’s impact on ideation, communication, and artistic workflows.

New methodologies must evaluate AI-assisted artistic creation holistically. He et al. (2023)\cite{he2023searching} propose co-creativity as a joint search process, offering insights into human-AI dynamics. Ethical concerns also arise, including copyright, ownership, and authenticity. Epstein et al. (2023)\cite{epstein2023art} call for interdisciplinary research into AI’s cultural and economic impact.

As AI-assisted art evolves, evaluation models must adapt. Future research should examine real-time, personalized experiences and how AI reshapes social interactions in creative settings~\cite{suh2021ai}. Documenting the artist-AI collaboration process may become as important as assessing final works.

\end{document}